\documentclass{aa}

\usepackage{natbib}
\bibpunct{(}{)}{;}{a}{}{,} 

\usepackage{graphicx}   
\usepackage{txfonts}
\usepackage[colorlinks=true,allcolors=cyan]{hyperref}%

\newcommand{\MSUN}{{\rm M}_\odot}

\begin{document}

\titlerunning{X-ray emission around GAMA galaxies in eROSITA/eFEDs}
\title{The eROSITA Final Equatorial Depth Survey (eFEDS):} 
\subtitle{X-ray emission around star-forming and quiescent galaxies at $0.05<z<0.3$}

\authorrunning{Comparat et al.}
\author{
Johan Comparat\inst{1}\thanks{E-mail: comparat@mpe.mpg.de}, 
Nhut Truong\inst{2}, 
Andrea Merloni\inst{1},        
Annalisa Pillepich\inst{2}, 
Gabriele Ponti\inst{3,1}, 
Simon Driver\inst{4,}\inst{5}, 
Sabine Bellstedt\inst{4},
Joe Liske\inst{6}, 
James Aird\inst{7,8},
Marcus Br\"uggen\inst{6},
Esra Bulbul\inst{1}, 
Luke Davies\inst{4},
Justo Antonio González Villalba\inst{1}, 
Antonis Georgakakis\inst{9},
Frank Haberl\inst{1}, 
Teng Liu\inst{1}, 
Chandreyee Maitra\inst{1},
Kirpal Nandra\inst{1}, 
Paola Popesso\inst{10}, 
Peter Predehl\inst{1}, 
Aaron Robotham\inst{4},
Mara Salvato\inst{1}, 
Jessica E. Thorne\inst{4},
Yi Zhang\inst{1}
}

 \institute{
 Max-Planck-Institut f\"{u}r extraterrestrische Physik (MPE), Gie{\ss}enbachstra{\ss}e 1, D-85748 Garching bei M\"unchen, Germany
 \and Max-Planck-Institut f\"{u}r Astronomie, K\"{o}nigstuhl 17, D-69117 Heidelberg, Germany
 \and INAF-Osservatorio Astronomico di Brera, Via E. Bianchi 46, I-23807 Merate (LC), Italy
 \and International Centre for Radio Astronomy Research (ICRAR), University of Western Australia, Perth, Western Australia, Australia; 
 \and International Space Centre (ISC), University of Western Australia, Perth, Western Australia, Australia
 \and Hamburger Sternwarte, University of Hamburg, Gojenbergsweg 112, D-21029 Hamburg, Germany
 \and Institute for Astronomy, Royal Observatory, University of Edinburgh, Edinburgh EH9 3HJ, UK
 \and School of Physics \& Astronomy, University of Leicester, University Road, Leicester LE1 7RJ, UK
 \and Institute for Astronomy and Astrophysics, National Observatory of Athens, V. Paulou \& I. Metaxa, 11532, Greece
 \and European Southern Observatory, D-85748 Garching bei M\"unchen, Germany
}
    
\date{\today}

\abstract
{}
{The circumgalactic medium (CGM) plays an important role in galaxy evolution as the main interface between the star-forming body of galaxies and the surrounding cosmic network of in- and out-flowing matter. In this work, we aim to characterize the hot phase of the CGM in a large sample of galaxies using recent soft-X-ray observations made by \textit{SRG}/eROSITA.}
{We stack X-ray events from the `eROSITA Final Equatorial Depth Survey' (eFEDS) around central galaxies in the 9hr field of the `GAlaxy and Mass Assembly' (GAMA) survey to construct radially projected X-ray luminosity profiles in the 0.5--2 keV rest frame energy band as a function of their stellar mass and specific star formation rate. We consider samples of quiescent (star-forming) galaxies in the stellar mass range $2\times 10^{10}$ -- $10^{12}$ M$_\odot$ ($3\times 10^{9}$ -- $6\times10^{11}$ M$_\odot$).} 
{
For quiescent galaxies, the X-ray profiles are clearly extended throughout the available mass range; however,
the measured profile is likely biased high because of projection effects, as these galaxies tend to live in dense and hot environments. 
For the most massive star-forming samples ($\geq10^{11}$ M$_\odot$), there is a hint of detection of extended emission. 
On the other hand, for star-forming galaxies with $< 10^{11}$ M$_\odot$ the X-ray stacked profiles are compatible with unresolved sources and are consistent with the expected emission from faint active galactic nuclei (AGN) and X-ray binaries.
We measure for the first time the mean relation between average X-ray luminosity and stellar mass separately for quiescent and star-forming galaxies. 
We find that the relation is different for the two galaxy populations: high-mass ($\geq 10^{11}$ M$_\odot$) star-forming or quiescent galaxies follow the expected scaling of virialized hot haloes, while lower mass star-forming galaxies show a less prominent luminosity and a weaker dependence on stellar mass consistent with empirical models of the population of weak AGN. 
When comparing our results with state-of-the-art numerical simulations (IllustrisTNG and EAGLE), we find overall consistency on the average emission on large ($>80$ kpc) scales at masses $\geq 10^{11}$ M$_\odot$, but disagreement on the small scales, where brighter-than-observed compact cores are predicted. 
The simulations also do not predict the clear differentiation that we observe between quiescent and star-forming galaxies in our samples.}
{This is a stepping stone towards a more profound understanding of the hot phase of the CGM, which holds a key role in the regulation of star formation. 
Future analysis using eROSITA all-sky survey data, combined with future generation galaxy evolution surveys, shall provide much enhanced quantitative measurements and mapping of the CGM and its hot phase(s).} 

\maketitle

%
%
%
%
\section{Introduction}
\label{sec:intro}

\begin{figure*}[h!]
    \centering
\includegraphics[width=1.9\columnwidth ]{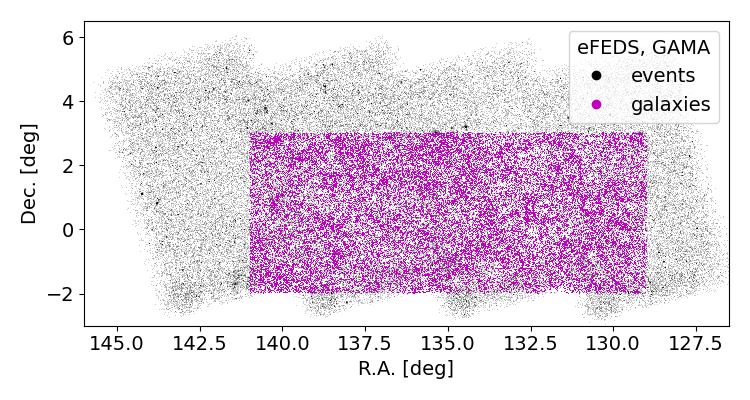}

\includegraphics[width=.62\columnwidth ]{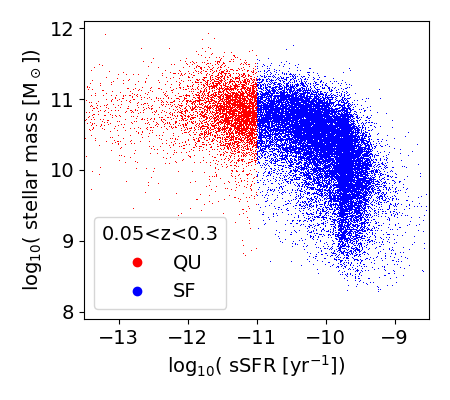}
\includegraphics[width=.62\columnwidth ]{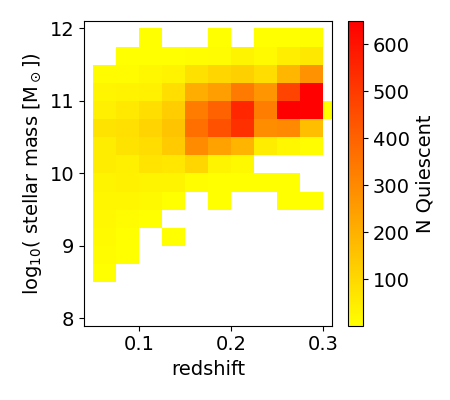}
\includegraphics[width=.62\columnwidth ]{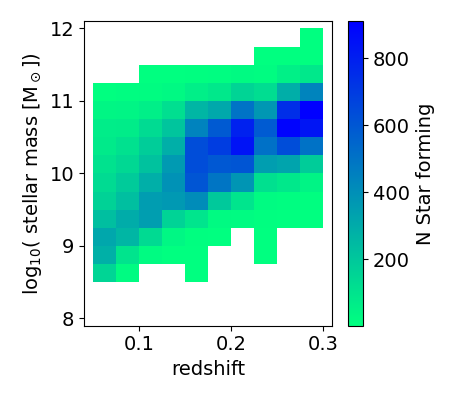}
    \caption{
    Central galaxies used in this work. 
    \textbf{Top.} Sky coverage of eFEDS data (gray points) and GAMA galaxies (magenta). 
    \textbf{Bottom panels.} 
    Split between star-forming and quiescent galaxies  (leftmost panel) formed by a boundary at $\log_{10}(sSFR)=-11$.
    Stellar mass ($\log_{10}(M/M_\odot)$) vs. redshift {2D histograms (redshift bins have a width of 0.025 and stellar mass bins have a width of 0.25 dex) for the set of quiescent (middle panel) and star-forming (right panel)} galaxies available in the GAMA 9hr field.} 
 \label{fig:GAMA:data}
\end{figure*}

A precise description of the different phases of the cosmic gas, from the intergalactic (IGM) to the circumgalactic medium (CGM) around galaxies, is the missing piece in the puzzle of the baryon budget in the Universe and currently prevents us from obtaining a complete and accurate description  \citep{tumlinson17, Driver2021NatAs...5..852D}. 
The physical properties of the warm-hot component of the intergalactic and halo gas, supposedly comprising 40\% of the baryons, have been vastly unconstrained until now. 
In this regard, $L^*$ galaxies hosted by $10^{12}\,\MSUN$ haloes, where most of the stars and metals are formed \citep{Moster2018MNRAS.477.1822M, Behroozi2019MNRAS.488.3143B}, are of great interest. 
Soft-X-ray observations represent a particularly useful tool for investigating the properties of the halo gas around galaxies, as the warm-hot phase itself should reach virial temperatures such that most of the emission emerges in this band \citep{fukugita04,Fukugita06}. 

Hot, X-ray-emitting haloes have been observed around {individual} or small samples of galaxies in the past. 
This has been achieved for mostly early-type massive galaxies \citep[e.g.,][]{goulding16,bregman2018}. 
The detection of X-ray-emitting atmospheres around star-forming disk galaxies is rare and limited to small samples of galaxies more massive than $\sim10^{11}\MSUN$ \citep[e.g.,][]{bogdan13a, bogdan13b, bogdan17, anderson16, li16}.

\citet{anderson15} stacked X-ray photons from the ROSAT all-sky survey around 250 000 massive galaxies from the Sloan Digital Sky Survey that are central within their dark-matter haloes. 
They reported a strong correlation between the mean X-ray luminosity of the volume-filling gas in the CGM (i.e., in the range $(0.15-1)\times R_{\rm 500c}$\footnote{$R_{\rm 500c}$ is the radius at which the density of the dark matter halo is 500 times the critical density of the Universe at the redshift of the system.}) and the galaxy stellar mass in the stellar mass range $\log M^*/\MSUN = 10.8-12$. However, because of the limited spatial resolution and the bright flux limit of ROSAT, it was not possible to firmly detect an X-ray emission signal from the CGM of Milky Way-mass (MW) galaxies and smaller ($\lesssim10^{10.7}\,\MSUN$). 

From a theoretical perspective, the extended soft-X-ray emission from the gaseous atmospheres of massive haloes has been predicted by galaxy formation models in the full cosmological context since the analytical models of \citet{White&Frenk1991}. This has been confirmed, albeit with overall lower luminosity than previously expected, by the results of cosmological hydrodynamical simulations, also around disk and MW-mass galaxies \citep[e.g.,][with the GIMIC, Illustris, and EAGLE simulations, respectively]{crain10, bogdan2015, kelly2021}. These works have shown that such X-ray emission is sensitive to the unfolding and the interplay of star formation, feedback, and cooling processes, which can simultaneously remove low-entropy gas by star formation, heat up the halo gas via energy injections, and redistribute the gas via outflows, making gaseous haloes dilute.
Recently, \citet{Truong20} and \citet{Oppenheimer2020} showed that the state-of-the-art cosmological galaxy simulations IllustrisTNG and EAGLE both predict an X-ray luminosity dichotomy at $z\sim0$: at the transitional stellar mass regime of $10^{10.5-11}\,\MSUN$, simulated star-forming galaxies exhibit somewhat {higher} soft-X-ray luminosity from the volume-filling gas within and around them than quiescent galaxies of the same mass, all the way out to galactocentric distances of $\sim$200 kpc. Despite the differences in feedback processes implemented therein, this has been shown to be a direct manifestation of the quenching mechanism in the simulations, with the reduction of the gas mass within the haloes being due to super massive black hole (SMBH)-driven outflows. Clearly, robust statistical constraints on the properties of the X-ray-emitting gas from large samples of galaxies in this transitional mass regime may hold the key to further improving our understanding of the complex physical processes shaping the gaseous atmosphere of their haloes.

In this article, we bridge the gap between past studies by attempting to measure the faint, extended X-ray emission ---that is, the so-called hot phase of the CGM--- surrounding central galaxies over a wide stellar mass range ($\approx10^{9.6-11.8}\,\MSUN$) by stacking \textit{SRG}/eROSITA data, taking advantage of its high sensitivity in soft X-rays, moderate spatial resolution, large grasp, and stable background \citep{Predehl2021}.
To this end, we use data from the eROSITA performance verification eFEDS field \citep{Brunner2021arXiv210614517B}, a 140 deg$^2$ survey that partly overlaps with a highly complete spectroscopic sample of low-redshift ($0.05<z<0.3$) galaxies \citep[GAMA, Sloan Digital Sky Survey (SDSS),][]{Liske2015MNRAS.452.2087L, Ahumada2020ApJS..249....3A}. 
We can therefore stack X-ray data around galaxies in different bins of stellar mass, and distinguish between star-forming and quiescent galaxies. 

A brief description of the data used is presented in Sect. \ref{sec:data}. 
The method, close to that adopted by \citet{anderson15}, is described in detail in Sect. \ref{sec:method}. 
In Sect. \ref{sec:result}, we discuss the measurements obtained, while in Sect. \ref{sec:simulation} we compare the measurements with hydrodynamical simulations. 
We discuss the possible implications of our results for galaxy evolution in Sect. \ref{sec:discussion}. 

%
%
%
%
\section{Data}
\label{sec:data}

\subsection{eROSITA eFEDS data}
\label{sec:masking}
We use for this work the public Early Data Release (EDR) eROSITA event file of the eFEDS field \citep{Brunner2021arXiv210614517B}\footnote{\url{https://erosita.mpe.mpg.de/edr}}. It contains about 11 million events (X-ray photons) detected by eROSITA over the 140 deg$^2$ area of the eFEDS Performance Verification survey.
Each photon is assigned an exposure time using the vignetting-corrected exposure map. 
Photons close to detected sources in the source catalog are flagged (see details in Sect. \ref{subsec:masking}). 
These sources are cataloged as point-like or extended based on their X-ray morphology \citep{Brunner2021arXiv210614517B}, and 
they are further classified (e.g., galactic, active galactic nuclei (AGN), individual galaxies at redshift $z<0.05$, galaxy group, and cluster) using multi-wavelength information \citep{Salvato2021arXiv210614520S, Vulic2021arXiv210614526V,  LiuAng2021arXiv210614518L, Liu2021arXiv210614522L, Bulbul2021arXiv211009544B}. 

\subsection{GAMA galaxy catalog}
\label{sec:GAMA}

{The Galaxy And Mass Assembly (GAMA) survey and its fourth data release provides spectroscopic redshift for more than 95\% of the galaxies brighter than $\sim$19.80 in the r-band \citep{Liske2015MNRAS.452.2087L,Driver2022MNRAS.tmp..552D}. 
The released data and galaxy catalog are described at length by \citet{Driver2022MNRAS.tmp..552D}\footnote{\url{http://www.gama-survey.org/dr4/}}. 
For each galaxy, using multi-wavelength observations covering the far-ultraviolet (FUV;\ $\sim$1500$\AA$) to the far-infrared (FIR; $\sim$500$\mu$m), GAMA constructed wide spectral energy distribution (SED) and adjusted the parameters of the stellar population constituting these galaxies \citep{Robotham2020MNRAS.495..905R,Bellstedt2020MNRAS.498.5581B,Bellstedt2021MNRAS.503.3309B}.
Of the GAMA fields, the 9hr field overlaps with the eFEDS observations; see Fig. \ref{fig:GAMA:data}. In that field, for galaxies brighter than $r < 19.8$, the spectroscopic completeness is 98.67\%.}

{From the 9hr field (SpectCatv27 and NQ$>2$), we retrieve about 40 000 galaxies with a spectroscopic redshift in the range 0.05$\leq z\leq $0.3 and with measured stellar mass and specific star formation rate (sSFR) from \citet{Bellstedt2020MNRAS.498.5581B,Bellstedt2021MNRAS.503.3309B}.
These are derived using SED fitting \citep{Robotham2020MNRAS.495..905R}. }
They adopt a Chabrier initial mass function (IMF) and SFRs are averaged over 100 Myr.
Stellar masses are output in units of solar mass ($M_\odot$) and sSFR per year (yr$^{-1}$). 
{The stellar mass function and cosmic SFR density are accurate and in excellent agreement with the literature \citep{Bellstedt2020MNRAS.498.5581B,Koushan2021MNRAS.503.2033K,Driver2022MNRAS.tmp..552D}.}

{The GAMA 9hr galaxy sample only covers a fraction of the eFEDS area: 60 deg$^2$ out of 140 deg$^2$; see top panel of Fig. \ref{fig:GAMA:data}. Necessarily, we limit our X-ray analysis to this 60 deg$^2$ area. The exact selection of the different galaxy subsamples adopted in our analysis is detailed in Sect. \ref{subsec:centrals}.} 

%
%
%
%
\section{Method}
\label{sec:method}

By stacking large data sets, the noise decreases and features with weak signal may be unveiled \citep[e.g.,][]{Zhu2015ApJ...815...48Z, Comparat2020AA...636A..97C, Wolf2020MNRAS.492.3580W}. 
To do so, we stack X-ray events around central galaxies with known spectroscopic redshift (Sect. \ref{subsec:centrals}), after masking detected X-ray sources (Sect. \ref{subsec:masking}).

We keep track of the angular distance between the detected X-ray events and the galaxies (and their redshifts) to build radial profiles. 
We record the event energies to build X-ray spectra. 
We obtain a data cube where angular coordinates are converted to proper distance (angle averaged) radii, and the energy (or wavelength) vector is shifted to the galaxy redshift. The stacking procedure is described in detail in Sect. \ref{subsec:cube:stacking}.

In the measurement process, control on two systematic features is key: the emission of the background and the instrumental signatures. 
To simplify our analysis, and in light of the expected SED of the signal we are interested in, we focus here on the rest-frame energy range between 0.5 and 2 keV. 
In doing so, we only consider the energy range where the background emission is dominated by the well-understood diffuse emission of the Milky Way and by the Cosmic X-ray Background \citep[e.g.,][ \textcolor{cyan}{Ponti et al. in preparation}]{Predehl2021, Liu2021arXiv210614522L}, while the contribution of the particle (unvignetted) background is negligible. 
Moreover, we avoid the complex response of the lowest energy range (below $\sim$0.4 keV in the observed frame) where both detector noise and the effects of the light leak on the TM5 and TM7 \citep[see][]{Predehl2021} could introduce yet uncalibrated systematic effects. 
As in \citet{Brunner2021arXiv210614517B}, we select good events from nominal field
of view, exclude bad pixels, and keep events with PATTERN$\leq$15, which includes single, double, triple, and quadruple events. 
Also, given the relatively low signal-to-noise ratio (S/N) achieved from relatively small galaxy samples, we only focus our attention on broad-band photometric measurements. Work is ongoing on the calibration of the low-energy response of eROSITA, and future works will explore the possibility of stacked spectral analysis also in the 0.15--0.4 keV observed-frame energy range.

We apply a bootstrap procedure to reliably estimate the mean expected background and its variance (Sect. \ref{subsec:randoms}). 
Finally, we estimate the spatial extent of the point-source profile using an empirical point spread function (PSF) model based on the detected point sources in the same eFEDS field, as we describe in Sect. \ref{subsec:equiv:psf}. 

\begin{table}
\centering
    \caption{
    Samples considered in the analysis.}
    \label{tab:gama:data}
    \begin{tabular}{r | l l r l l }
    \hline \hline
    Sample & \multicolumn{2}{c}{stellar mass} & $N_g$ & \multicolumn{2}{c}{Average}  \\ 
    name & min & max & & $\bar{M}$  & $\bar{z}$ \\ 
    \hline
ALL\_M10.7 & 10.4 & 11.0 & 16142 & 10.7 & 0.22 \\
\hline
& \multicolumn{4}{c}{Fixed stellar mass selection} \\
\hline
QU\_M10.7 & 10.0 & 11.0 & 7267 & 10.72 & 0.2 \\
SF\_M10.7 & 10.4 & 11.0 & 9846 & 10.66 & 0.23 \\
\hline
& \multicolumn{4}{c}{Quiescent galaxies} \\
\hline
QU\_M11.71 & 11.616 & 11.973 & 50 & 11.71 & 0.27 \\
QU\_M11.58 & 11.556 & 11.616 & 50 & 11.58 & 0.27 \\
QU\_M11.54 & 11.523 & 11.556 & 50 & 11.54 & 0.26 \\
\hline
QU\_M11.35 & 11.306 & 11.41 & 400 & 11.35 & 0.25 \\
\hline
QU\_M11.2 & 11.138 & 11.269 & 1002 & 11.2 & 0.25 \\
\hline
QU\_M11.04 & 10.961 & 11.138 & 2000 & 11.04 & 0.24 \\
QU\_M10.88 & 10.806 & 10.961 & 1999 & 10.88 & 0.23 \\
QU\_M10.73 & 10.641 & 10.806 & 1999 & 10.73 & 0.21 \\
QU\_M10.53 & 10.362 & 10.641 & 2000 & 10.53 & 0.19 \\
\hline
& \multicolumn{4}{c}{star-forming galaxies} \\
\hline
SF\_M11.25 & 11.17 & 11.754 & 400 & 11.25 & 0.27 \\
SF\_M11.12 & 11.079 & 11.17 & 400 & 11.12 & 0.27 \\
SF\_M11.05 & 11.027 & 11.079 & 401 & 11.05 & 0.26 \\
\hline
SF\_M10.99 & 10.943 & 11.051 & 1000 & 10.99 & 0.26 \\
SF\_M10.9 & 10.861 & 10.943 & 1001 & 10.9 & 0.25 \\
\hline
SF\_M10.86 & 10.795 & 10.943 & 2000 & 10.86 & 0.25 \\
SF\_M10.74 & 10.68 & 10.795 & 2000 & 10.74 & 0.24 \\
SF\_M10.63 & 10.574 & 10.68 & 2002 & 10.63 & 0.23 \\
SF\_M10.52 & 10.467 & 10.574 & 1998 & 10.52 & 0.22 \\
SF\_M10.41 & 10.358 & 10.467 & 2001 & 10.41 & 0.21 \\
SF\_M10.3 & 10.241 & 10.358 & 2000 & 10.3 & 0.2 \\
SF\_M10.18 & 10.108 & 10.241 & 2001 & 10.18 & 0.19 \\
SF\_M10.03 & 9.947 & 10.108 & 1998 & 10.03 & 0.18 \\
SF\_M9.86 & 9.761 & 9.947 & 2000 & 9.86 & 0.16 \\
SF\_M9.64 & 9.491 & 9.761 & 2000 & 9.64 & 0.14 \\
\hline
\end{tabular}
\tablefoot{
SF stands for star-forming and QU for quiescent. 
We report the number of galaxies present in each sample ($N_g$) and their average properties: stellar mass ($\bar{M}$) and redshift ($\bar{z}$).
}
\end{table}

\subsection{Selecting galaxies}
\label{subsec:centrals}

We select central galaxies, most massive within their host dark matter halo, similarly to \citet{planck13}. 
For each GAMA galaxy, we infer its host halo mass and corresponding virial radius with the stellar-to-halo-mass relation from \citet{moster13}. 
If a galaxy lies within twice the virial radius of a galaxy of higher stellar mass, it is considered a satellite and is removed from the sample. The choice of two times the virial radius is a conservative one in order to account for the scatter in the stellar-mass to halo-mass relation. 
We treat the X-ray-detected eFEDS clusters \citep{LiuAng2021arXiv210614518L}  separately, for which we have individual measurements of R$_{500c}$ \citep{Bahar2021arXiv211009534B}. For them,
we only remove satellites falling within one virial radius, taken as R$_{500c}$/0.7. 
After this filtering, $\sim10\%$ of the galaxies are removed, and we obtain a sample of 35\,521 central galaxies. 
Thanks to the high completeness ($\sim 98\%$) of the GAMA sample, the sample of central galaxies should also be highly complete. 
We discuss limitations due to our sample definition in Sect. \ref{subsec:disc:projection}

We use the reported stellar masses and sSFRs to create subsamples of the galaxy population. 
In order to examine trends, we split the population into star-forming and quiescent galaxies assuming a boundary fixed at $\log_{10}(sSFR)=-11$ \citep[see discussion by ][]{Davies2019MNRAS.483.1881D, Thorne2021MNRAS.505..540T}.
For this study, in which we stack around a large number of galaxies, the exact boundary definition should have a minor impact. 
Figure \ref{fig:GAMA:data} shows the distribution of galaxies in the redshift range of interest in the mass--sSFR plane. 

To compare the star-forming and quiescent samples at fixed stellar mass, we first adopt a stellar mass selection to obtain two samples with the same mean stellar mass, different sSFR, and a similar total number of galaxies. 
By taking objects within $10<\log_{10} M^*<11$ for the quiescent and $10.4<\log_{10} M^*<11$ for the star-forming galaxies, we obtain a mean stellar mass of $5\times10^{10}M_\odot$  for both, with a set of 7\,267 and 9\,846 galaxies, respectively. 

Each population, star-forming or quiescent, is then split in a number of nonoverlapping stellar mass subsamples (see Table \ref{tab:gama:data}). 
As stellar mass should correlate with X-ray luminosity \citep{anderson15}, in order to obtain a similar S/N from the various subsamples, fewer galaxies are needed at higher mass than at lower mass. 
We therefore create subsamples of $\sim$2000 at the low-mass end, then 1000, 400, and finally 50 galaxies at the high-mass end. 
Table \ref{tab:gama:data} details the exact number of galaxies present in each subclass. 
There, we also report the mean redshift and mean stellar mass for each subsample defined in this way. 

\subsection{Masking approaches and possible sources of contamination}
\label{subsec:masking}
As we are looking for faint diffuse emission, it is vital to remove (mask out) as many sources of contamination  as possible, that is, those produced by unresolved emission from compact sources within galaxies.
In this work, we investigate four possible masking schemes (see Table~\ref{tab:mask:length}):
\begin{enumerate}
    \item[(i)] `ALL' mask: all detected X-ray sources are masked;
    \item[(ii)] `M1' mask: all detected sources are masked except for those associated with a cluster or a group in the same redshift range as the GAMA galaxies, as identified by \citet{LiuAng2021arXiv210614518L} or by \citet{Bulbul2021arXiv211009544B}, taking \texttt{CLUSTER\_CLASS} = 4 or 5,  \citep[see][]{Salvato2021arXiv210614520S};
    \item[(iii)] `M2' mask: all detected sources are masked except for point sources in the redshift range of interest associated by \citet{Salvato2021arXiv210614520S} to a GAMA galaxy;
    \item[(iv)] `M3' mask: all detected sources are masked except for those unmasked by the M1 or M2 mask. The signal obtained is to be interpreted as the sum of all emitting entities: AGN, X-ray binaries (XRBs), and hot gas augmented by systematic projection effects. 
\end{enumerate} 

The masking radius for each detected source (with a detection likelihood of greater than 6) is its radius of maximum S/N, as determined while extracting the X-ray spectrum of each source \citep{Liu2021arXiv210614522L}, augmented by 40\%. 
By doing so, we make sure there is no remaining correlation between the set of events outside of the mask and the source catalog (Comparat et al. in prep.). 
The optimal masking radius, derived with limited statistics on eFEDS data, suffers from uncertainties; augmenting the masking radius by 30\% (or 50\%) is also reasonable and corresponds to masking 2\%\ fewer (or more) events, as shown in Table \ref{tab:mask:length}. 
This uncertainty on the total number of events directly impacts the normalization of the profiles estimated.
To account for this, we add a systematic 2\% uncertainty on the background luminosity density (our normalization); see Sect. \ref{subsec:randoms}. 

\begin{table}
    \centering
    \caption{Fraction of masked events \label{tab:mask:length}}    
    \begin{tabular}{c c c c c c}
\hline \hline
mask & \multicolumn{2}{c}{fraction of masked events}  & masked area  \\
name & \multicolumn{2}{c}{\% augmentation of masking radius}  & fraction \\
 & 40 & (30, 50) & $a_{\rm mask}$ \\ \hline
ALL  & 0.2713 & (0.2565,  0.2863) &   0.069 & \\
M1   & 0.259  & (0.245,  0.2732) &   0.066 & \\
M2   & 0.2542 & (0.2394,  0.2696) &   0.065 & \\
M3   & 0.2414 & (0.2275,  0.256) &   0.061 & \\
\hline
    \end{tabular}
    \tablefoot{Fraction of masked events for a radius of maximum S/N augmented by 30\%\ to 50\%, where 
    40\% is the baseline used in the analysis. 
    Masks are ordered by decreasing masked fraction. 
    The percentage of area masked corresponding to the 40\% baseline is given in the last column.}
\end{table}

We use the sensitivity map to generate a catalog of random points following  \citet{Georgakakis2008MNRAS.388.1205G}. Armed with this, we estimate that masking all X-ray sources removes 27.13\% of the events (and 6.9\% of the area) in the 0.5--2 keV band. 
The least stringent mask, `M3', removes 24.14\% of the events (and 6.1\% of the area; see Table \ref{tab:mask:length}).

The baseline mask used in this study is the M1 mask: all sources are masked except for sources identified as galaxy groups or clusters at $0.05\leq z\leq0.3$. 
Indeed, masking these extended sources would bias low the X-ray profiles of high-mass galaxies. 
The other masks enable us to investigate systematic effects due to the masking procedure. In particular, we present a detailed comparison with the results from the M3 mask, which include the emission from all the GAMA sources detected by eROSITA as point sources.
{The set of GAMA galaxies matched to X-ray point sources sample both the unobscured and the obscured AGN loci. 
Their luminosity ranges from $5\times 10^{40}$ to $2\times 10^{44}$ erg s$^{-1}$ in the soft-X-ray band.} Below, we discuss in more detail the possible contamination due to faint, undetected AGN or XRBs, and the relationship with the alternative masking approaches.

\subsubsection{Expected AGN signal}
\label{subsec:agn}

X-ray emission from AGN is produced in a very compact (fraction of a milli-parsec) region close to the central SMBHs in the nuclei of galaxies, and thus represents a contamination to the CGM signal. To ease the interpretation of the stacked profiles, we would therefore ideally remove as many active galaxies from the sample as possible. 
However, given the moderate angular resolution of eROSITA, this step is far from straightforward. In order to assess our ability to remove AGN contaminants, we first discuss the completeness of GAMA towards X-ray AGN detected by eROSITA. 

Within the eFEDS X-ray point-source catalog, considering all those counterparts in the GAMA 9hr. field and in the redshift range $0.05<z<0.3$, using either spectroscopic or high-quality photometric redshifts \citep[see][]{Salvato2021arXiv210614520S}, we obtain 619 X-ray sources. Of these, 474 (76.6\%) are matched within 2\arcsec~ to a galaxy present in the GAMA catalog.
When limiting the X-ray catalog to sources with an LS8 magnitude r$<19.8$ (19), similar to the magnitude limit used in GAMA, 88.8\% (90.2\%) are matched to GAMA galaxies.
This implies that, at the magnitude limit of GAMA, the galaxy catalog is nearly complete in terms of counterparts of the bright X-ray point-source population detected in eFEDS. In turn, this is consistent with the known spectroscopic completeness achieved by GAMA at these magnitude limits, and with the GAMA target selection \citep{Baldry2010MNRAS.404...86B}, which uses a combination of criteria to exclude stars, while keeping compact galaxies and quasi-stellar objects (QSOs).
The remaining unmatched X-ray sources are typically fainter in the optical than the GAMA limit, and are always masked out in the stacks. 

The point-source X-ray flux limit of eFEDS, of namely $\sim 6.5\times 10^{-15}$ erg s$^{-1}$ cm$^{-2}$ in the 0.5--2 keV band \citep{Brunner2021arXiv210614517B},  corresponds to a rest-frame luminosity of between about 5$\times 10^{40}$ and 2$\times 10^{42}$ erg s$^{-1}$  at the redshift of the GAMA sources we are interested in. {We therefore detect essentially all X-ray-bright AGN among GAMA galaxies; removing photons around all detected point-like X-ray sources (M1 mask) therefore removes the contamination from all the bright AGN from the sample.}
However, within the GAMA galaxy catalogs, a fraction of galaxies are expected to host fainter AGN, which remain undetected given the eROSITA/eFEDS flux limit.
Aird et al. (in preparation) study the point-source emission emerging from GAMA galaxy stacks (as a function of stellar mass and redshift) to determine the faint end of the AGN X-ray luminosity function.
These authors measure and model the average luminosity and the fraction of galaxies hosting an X-ray AGN. 
For a stellar mass ($\log_{10}(M^*/[M_\odot])$) of 9.75 (10.75, 11.75), they find an average luminosity of $\log_{10}(L_X/{\rm[erg\; s^{-1}]}) \approx$ 40, (41, 42) and an occupation fraction of 0.1\% (1\%, 10\%). We further discuss AGN contamination and compare these figures with our observations in Sect. \ref{subsubsec:agn_cont_discussion}.

\subsubsection{Expected X-ray binary signal}
\label{subsec:xrb}

X-ray binaries, the end points of stellar evolution, are known contributors to the total X-ray luminosity of a galaxy \citep{Tauris2006csxs.book..623T}. They are typically spatially distributed following the stellar light, and therefore their emission would be unresolved by eROSITA at the redshift of interest here.

We evaluate the possible contribution from these (unresolved) XRBs by taking advantage of the known scaling relation between their total X-ray luminosity and their host galaxy properties.
In particular, in order to predict the X-ray luminosity   emitted by each galaxy and attributable to these sources, we use an analytical model based on  
\citet{Lehmer2016ApJ...825....7L} and \citet{Aird2017MNRAS.465.3390A}. 
These authors measured the dependence of the total XRB luminosity (in the 2--10 keV energy band) on redshift, galaxy stellar mass, and SFR. 
To make sure our prediction is conservative, we use the  \citet{Aird2017MNRAS.465.3390A} model, which produces a 10\%-20\% brighter XRB luminosity for a given galaxy property. 
We use their "model 5" with parameters given in their Table 3 to predict the X-ray luminosity in the band 2-10 keV. 
We propagate the uncertainties from this latter table to the prediction. 
We then convert (multiplication by 0.56) the luminosity to the 0.5--2 keV band assuming an absorbed (with n$_H$ column density set at mean value of the field $4\times10^{20}$ cm$^{-2}$) power law with a photon index of 1.8 \citep[as suggested by ][]{BasuZych2020MNRAS.498.1651B}. {We note here that all the X-ray-detected point-like sources in eFEDS} are significantly more luminous than predicted by the \citet{Aird2017MNRAS.465.3390A} model.

\subsection{Stacking procedure}
\label{subsec:cube:stacking}

We assume a flat $\Lambda$CDM cosmology with $H_0=67.74$ km s$^{-1}$ Mpc$^{-1}$, $\Omega_m(z=0)=0.3089$ \citep{Planck2016AA...594A..24P}.
Each galaxy is characterized by its position on the sky and its redshift, as well as properties of its stellar population (mass, sSFR). 
We denote a galaxy with the vector $\vec{G}$ defined as
\begin{equation}
\vec{G} = (G_{RA}, G_{Dec}, G_{z}, G_{M}, G_{sSFR}).
\end{equation}

For each galaxy, we retrieve all the events within the angle subtended by 3 Mpc at the galaxy redshift. 
We construct a "cube" of events surrounding each galaxy. 
For each event, we compute a rest-frame energy by multiplying the energy by one plus the galaxy redshift: $E_{rf}=E_{obs}\times(1+G_z)$. 
Therefore, each eROSITA event is characterized by the following vectors: its position on the sky (R.A., Dec.), its rest frame and observed energy, the corresponding galaxy redshift, the exposure time, the on-axis telescope effective area as a function of energy (ARF) at the observed energy, and the projected distance ($R_p$) in proper kiloparsecs (kpc) to the galaxy. 
We denote an event with the vector $\vec{E}$ defined as
\begin{equation}
\vec{E} = (\textrm{R.A.}, \textrm{Dec.}, E_{obs}, E_{rf}, G_{z}, t_{exp}, ARF(E_{obs}), R_p).
\end{equation}
The exposure times are obtained from the vignetted exposure map \citep{Brunner2021arXiv210614517B}. 
Using the (angular) projected distance induces projection effects which we discuss in Sect. \ref{subsec:disc:projection}.
We repeat this procedure with sets of random locations in the field, replacing the galaxy positions with randomly drawn positions in the same area of the sky, taking advantage of the relatively uniform exposure of the eFEDS field \citep{Brunner2021arXiv210614517B}. 

Finally, in order to derive accurate correction to the measured fluxes for masking and boundary effects (due to the reduction of projected area), we repeat the above procedure with another two sets of random events. 
A first set of random events uniformly samples the area covered in the GAMA field ($RE_{G}$). 
A second set of events uniformly samples the GAMA area and an additional 1.5 degree wide stripe around ($RE_{W}$). This allows us to account for boundary effects in the area normalization of the background counts (see Sect. \ref{subsec:randoms}).

We apply each selection defined in Table \ref{tab:gama:data} to the galaxy sample and to the random galaxy samples. 
We concatenate the event sets obtained.
For each galaxy sample, we obtain five cubes of events: galaxy events (data cube), random galaxy events (random cube), point-source galaxy events (point-source cube, detailed in Sect. \ref{subsec:equiv:psf}), galaxies-$RE_{G}$ cube, and galaxies-$RE_{W}$ cube.

Each event in any of the cubes is weighted by the following function:
\begin{equation}
w_i = \frac{A_{\rm corr}(r) \times 1.602177 \times 10^{-12} E_{rf}}{ARF(E_{obs})} \frac{ 4 \pi d^2_L(G_z) }{t_{exp} \times N_g},  
\end{equation}
where $N_g$ is the number of galaxies in a sample (given in Table \ref{tab:gama:data}). $A_{\rm corr}$ is the area-correction term, which accounts for both boundary effects and masks: 
\begin{equation}
    A_{\rm corr}(r) = 1 + \frac{ RE_{W}(r) } { RE_{G}(r) }  + a_{\rm mask},
\end{equation}
where $r$ is the proper projected separation in kpc. 
For the full sample, the correction $\frac{ RE_{W}(r) } { RE_{G}(r) }$ is 0.5\% at 100 kpc, 1\% at 300 kpc, 2.5\% at 1000 kpc, and 5 \% at 2 Mpc. 
For the M1 mask, $a_{\rm mask}=0.066$, and for other masks, values are given in Table \ref{tab:mask:length}.

A surface luminosity projected profile in erg s$^{-1}$ kpc$^{-2}$ {denoted $S_X$} is obtained from the weighted (using $w_i$) histogram of projected separations to the galaxies ($R_p$) divided by the area in kpc$^2$ covered by each histogram bin: $\pi (R^1_1 - R^2_0) A_{corr}$. 
Conversely, an X-ray spectrum, in erg s$^{-1}$ keV$^{-1}$ in a given projected radial aperture, is obtained with the weighted sum of the energies ($E_{rf}$) of all events in a given energy bin divided by the width of that bin (in keV).

\subsection{Background estimation and its uncertainties}
\label{subsec:randoms}
The projected radial profiles and integrated spectra obtained with the random cubes represent the null hypothesis of no signal, and are used to assess the robustness of any possible detection from the stacking samples. 
We repeat this measurement process at random positions 20 times (using different random points each time). 
From the outer shells of the radial background profiles (500 kpc to 3000 Mpc) of the 20 realizations, we estimate the mean background luminosity.
It takes values of around 1.1$\times 10^{37}$ erg s$^{-1}$ kpc$^{-2}$. 
For each galaxy sample, the background value obtained is different; indeed the total area covered and the masked area both vary from sample to sample.

The uncertainty due to the source-masking procedure (see Sect. \ref{subsec:masking}) suggests that the total number of events is subject to a residual 2\% systematic uncertainty  at most. 
To be conservative, we therefore add a 2\% systematic uncertainty to the mean value of the background: $\sigma^{BG}=0.02$  at all scales and energies.
The uncertainty on the galaxy stack count rates follows a Poisson distribution: $\sigma^{GAL}=1/\sqrt{N^{GAL}}$. 
The uncertainty on the final background-subtracted measurement (i.e., galaxy minus background) is the quadratic sum of the two uncertainties: $\sqrt{(\sigma^{GAL})^2+(\sigma^{BG})^2}=\sqrt{1/N^{GAL}+0.02^2}$. 

Additional sources of uncertainty may arise from the use of inaccurate redshifts, source positions, or photon energies. 
Uncertainties on centering and positions could artificially cause the PSF to broaden;
uncertainties on energies could cause spectral distortions; redshift uncertainties would cause both spectral and spatial distortions.
As we use spectroscopic redshifts, we consider the uncertainty coming from those to be negligible;
however, one could imagine that a few catastrophically incorrect redshifts are included. 
If, for example, these are additionally located in bright clusters (illustrative purpose), we would incorrectly convert arcminutes to kpc, which could cause profiles to either be more concentrated (redshift is lower than true redshift) or diluted (redshift measured is higher than true redshift). 
The exact quantification of a possible systematic error arising from catastrophic redshift, incorrect source positions, or photon energies is left for future studies. 

\begin{figure}
    \centering
    \includegraphics[width=0.9\columnwidth]{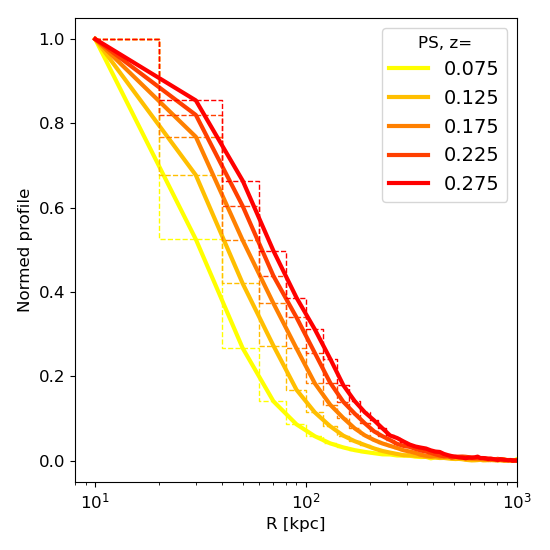}
    \caption{eROSITA normalized point-source profiles as a function of redshift in bins of 20kpc. Each curve represents the average point-source profile in a redshift bin of width of 0.05: 0.05-0.1, 0.1-0.15, 0.15-0.2, 0.2-0.25, 0.25-0.3. The labeled number gives the mean redshift of the bin. At the mean redshift of the sample ($z\sim 0.2$), the half width at half maximum (HWHM) of the empirical PSF corresponds to about 60 kpc.}
    \label{fig:psf:vs:redshift}
\end{figure}

\begin{figure*}
    \centering
\includegraphics[width=.67\columnwidth ]{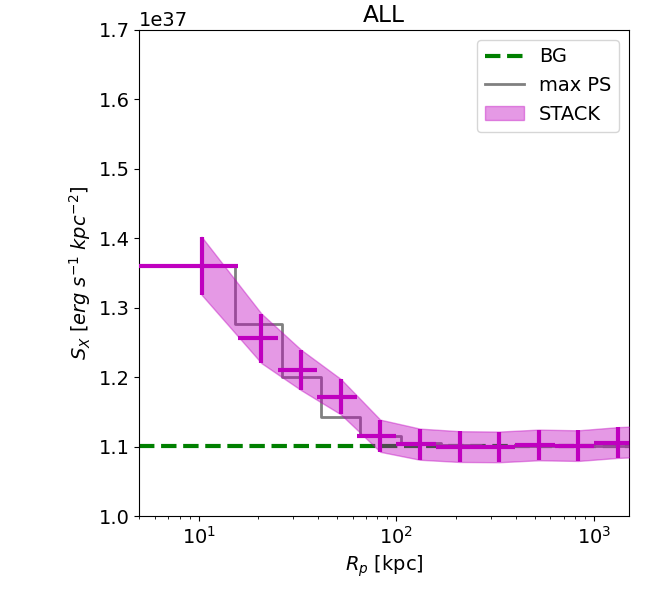}
\includegraphics[width=.67\columnwidth ]{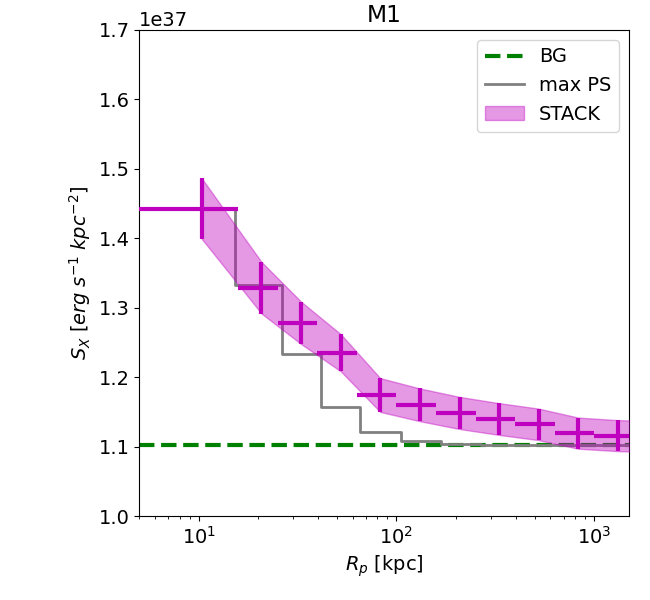}
\includegraphics[width=.67\columnwidth ]{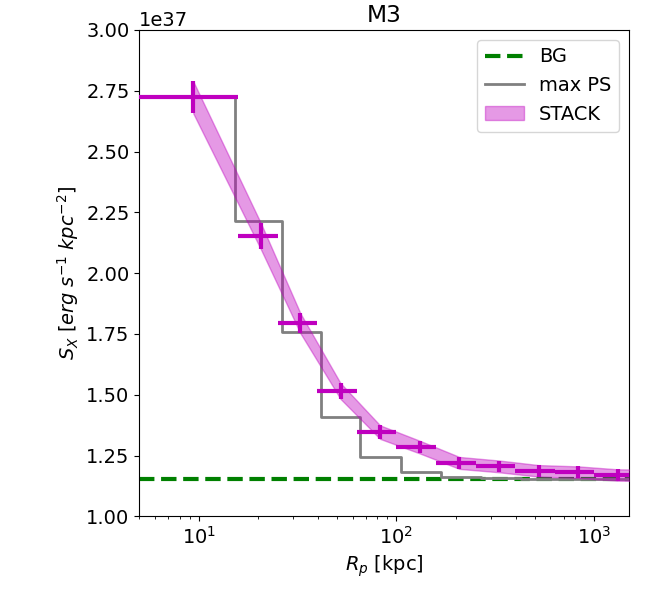}

\includegraphics[width=.67\columnwidth ]{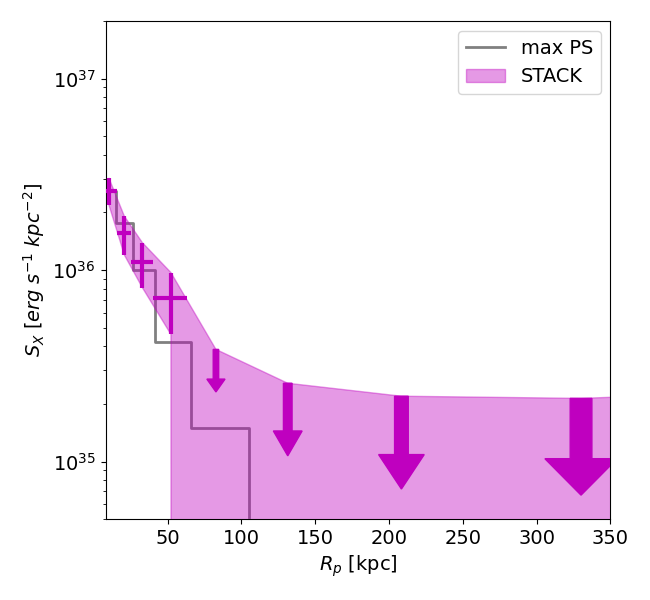}
\includegraphics[width=.67\columnwidth ]{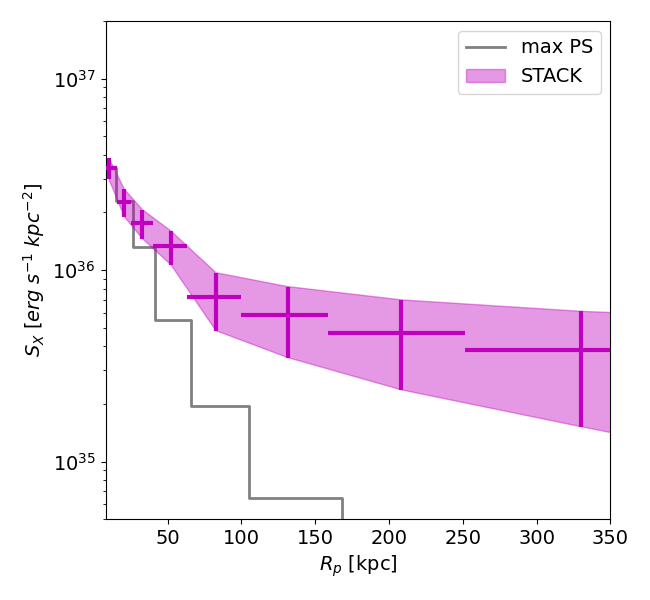}
\includegraphics[width=.67\columnwidth ]{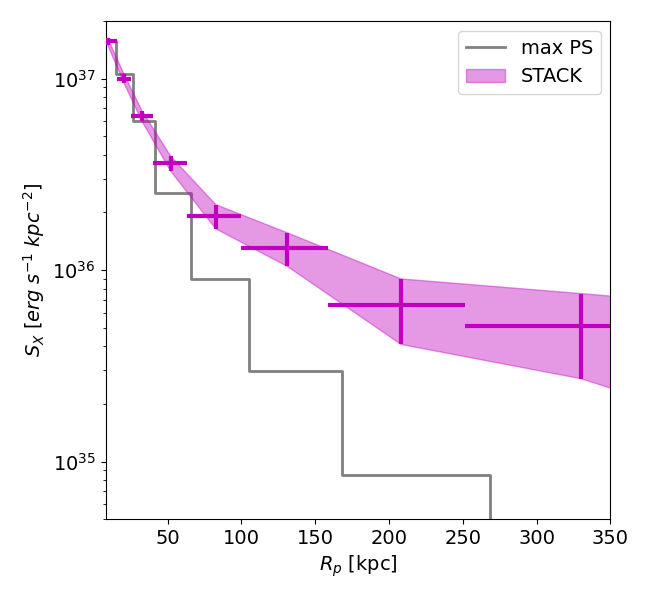}

    \caption{Measured X-ray radial projected luminosity profiles (0.5--2.0 keV rest-frame) for the ALL\_M10.7 GAMA central galaxy sample (`STACK', magenta crosses and shaded area). Each panel shows the result when a different mask is applied to the set of events: "ALL" (left), "M1" (middle), and "M3" (right). 
    We note the variation in the y-axis range in different panels. 
    The green dashed line represents the background level, estimated as discussed in Sect. \ref{subsec:randoms}.
    The profile shape expected if all sources stacked were point-like is shown with a gray line labeled "max PS". 
    The bottom series of panels shows the background-subtracted profiles with a linear radial scale extending to 300 kpc.}
    \label{fig:full:stack:measurement}
\end{figure*}

\subsection{Empirical point-source profile and validation against AGN}
\label{subsec:equiv:psf}
As we are interested in detecting extended CGM X-ray emission around galaxies, a key prerequisite is an accurate characterization of the eROSITA point spread function (PSF) and its convolution with the redshift distribution of the galaxies.

To obtain an empirical point-source profile for comparison to each galaxy sample, we repeat the procedure described in Sect. \ref{subsec:cube:stacking} with sets of detected X-ray point sources in eFEDS ("point-source cube").
{Each galaxy is artificially moved to the sky position of an X-ray point source. To do that, we replace the galaxy positions (on sky R.A., Dec.) with that} of extra-galactic point sources with moderately bright fluxes $10^{-14}<F_{\rm 0.2-2.3 keV}/[{\rm erg}\, {\rm cm}^{-2}\, {\rm s}^{-1}]<10^{-12}$ and {\tt ERO\_DET\_LIKE} $>20$, taken from the  \citet{Brunner2021arXiv210614517B} catalog. 

In doing so, we convolve the eROSITA PSF with the redshift distribution of the galaxy sample, and we obtain the shape of the radial profile expected if all sources were bright and point-like in the eROSITA images. 
Figure \ref{fig:psf:vs:redshift} shows how these empirical PSF profiles (in kpc) evolve  as a function of redshift. 
The higher the redshift, the broader the point-source profile. 
We do not stack beyond redshift 0.3 to avoid an overly wide PSF in kpc. 
At the mean redshift of the sample ($z\sim 0.2$), the HWHM of the empirical PSF corresponds to about 60 kpc.

We stress here that the purpose of this exercise is not to determine an accurate PSF profile for eROSITA \citep[see e.g.,][]{Churazov2020arXiv201211627C}, but rather to have a term of reference with which to assess the possible extended nature of the profiles measured around galaxies. 
The stacked profiles obtained here from the detected point sources are by construction clearly much brighter than the stacked galaxy profiles (see Sect. \ref{sec:result} below). To ease comparison, in each of the galaxy stacks we present below, we re-scale the convolved PSF profile to match the central value of the galaxy profile, creating a "maximal point source" (max PS) term of comparison\footnote{PSF profiles could be artificially broadened due to the clustering of the galaxies \citep{Popesso2012arXiv1211.4257P}. 
Complete simulations of the galaxy population and its X-ray emission would be needed over cosmological volumes to enable a quantitative assessment. 
Indeed, we need to generate a model to populate the full sky with X-ray-emitting galaxies together with their CGM \citep[possibly following simulations of the gas around galaxy clusters from][]{Comparat2020OJAp....3E..13C}, which is beyond the scope of this article. We defer the quantification of this effect  to a future study.}.

\subsection{Validation}
To validate the stacking procedure,
we apply it to known (eFEDS-detected) AGN (with measured spectroscopic redshift). 
We stack at the spectroscopic AGN redshift.
Its integrated luminosity (in erg s$^{-1}$) amounts to $\log_{10}(L_X) = 42.72 \pm 0.08$, while the mean luminosity of the same AGN set as determined by \citet{Liu2021arXiv210614522L} is $\log_{10}(L_X)=42.75$. 
The background-subtracted X-ray spectrum obtained is well fit by a power law with a photon index of 2.05$\pm0.05$, compatible with the mean slope of 2.02 determined on the same sample by \citet{Liu2021arXiv210614522L}. 
The projected luminosity profile and stacked spectra are therefore in very good agreement with the mean of the measurements made on individual AGN. 

\begin{table*}
    \centering    
    \caption{\label{tab:LX:80kpc:M1}Cylindrical projected X-ray luminosity using the M1 mask.}
    \begin{tabular}{c |c c c |c c c c c c c c c c}
\hline \hline
 sample &  $ \log_{10}(M_{vir}) $   & \multicolumn{2}{c|}{S/N} & \multicolumn{4}{c}{$L_{X}$ [$10^{40}$ erg s$^{-1}$]} \\
 name   & halo & 80 & 300 & XRB & max PS &  $R_p$<80\,{\rm kpc}& $R_p$<300\,{\rm kpc} \\
\hline
ALL\_M10.7 & 12.2 & 5.2 & 2.3 & 1.1 $^{+ 0.3 }_{- 0.2 }$ & 2.8 & 2.8 $\pm$ 0.5 & 15.4 $\pm$ 6.6 \\
\hline
& \multicolumn{4}{c}{Fixed stellar mass selection} \\
\hline
QU\_M10.7 & 12.2 & 7.3 & 4.3 & 0.6 $^{+ 0.2 }_{- 0.1 }$ & 3.1 & 4.0 $\pm$ 0.5 & 28.7 $\pm$ 6.6 \\
SF\_M10.7 & 12.2 & 3.6 & 0.9 & 1.4 $^{+ 0.4 }_{- 0.3 }$ & 2.6 & 1.9 $\pm$ 0.5 & 5.7 $\pm$ 6.7 \\
\hline
& \multicolumn{4}{c}{Quiescent galaxies} \\
\hline
QU\_M11.71 & 15.0 & 10.9 & 10.3 & 7.1 $^{+ 2.1 }_{- 1.6 }$ & 92.2 & 82.2 $\pm$ 7.6 & 440.7 $\pm$ 42.6 \\
QU\_M11.58 & 14.5 & 8.5 & 7.5 & 5.3 $^{+ 1.6 }_{- 1.2 }$ & 78.0 & 70.0 $\pm$ 8.2 & 314.3 $\pm$ 42.1 \\
QU\_M11.54 & 14.4 & 3.9 & 5.0 & 4.8 $^{+ 1.4 }_{- 1.1 }$ & 24.8 & 22.3 $\pm$ 5.7 & 174.1 $\pm$ 35.0 \\
\hline
QU\_M11.35 & 13.8 & 8.6 & 7.0 & 3.0 $^{+ 0.9 }_{- 0.7 }$ & 17.1 & 14.9 $\pm$ 1.7 & 87.5 $\pm$ 12.5 \\
\hline
QU\_M11.2 & 13.2 & 8.4 & 4.4 & 2.1 $^{+ 0.6 }_{- 0.5 }$ & 10.0 & 8.9 $\pm$ 1.1 & 40.5 $\pm$ 9.2 \\
\hline
QU\_M11.04 & 12.8 & 6.3 & 4.5 & 1.5 $^{+ 0.4 }_{- 0.3 }$ & 3.7 & 5.0 $\pm$ 0.8 & 36.7 $\pm$ 8.2 \\
QU\_M10.88 & 12.5 & 6.1 & 5.2 & 1.0 $^{+ 0.3 }_{- 0.2 }$ & 3.8 & 4.6 $\pm$ 0.8 & 41.5 $\pm$ 8.0 \\
QU\_M10.73 & 12.3 & 5.7 & 3.1 & 0.7 $^{+ 0.2 }_{- 0.1 }$ & 2.9 & 3.9 $\pm$ 0.7 & 22.4 $\pm$ 7.2 \\
QU\_M10.53 & 12.0 & 5.5 & 3.5 & 0.4 $^{+ 0.1 }_{- 0.1 }$ & 2.4 & 3.2 $\pm$ 0.6 & 22.5 $\pm$ 6.4 \\
\hline
& \multicolumn{4}{c}{star-forming galaxies} \\
\hline
SF\_M11.25 & 13.4 & 3.3 & 1.7 & 4.0 $^{+ 1.3 }_{- 1.0 }$ & 10.3 & 5.7 $\pm$ 1.7 & 21.8 $\pm$ 12.7 \\
SF\_M11.12 & 13.0 & 2.5 & 1.5 & 3.1 $^{+ 1.0 }_{- 0.7 }$ & 7.9 & 4.1 $\pm$ 1.6 & 18.8 $\pm$ 12.2 \\
SF\_M11.05 & 12.8 & 2.4 & 1.7 & 2.6 $^{+ 0.8 }_{- 0.6 }$ & 4.8 & 3.7 $\pm$ 1.6 & 20.6 $\pm$ 11.9 \\
\hline
SF\_M10.99 & 12.7 & 3.5 & 1.3 & 2.4 $^{+ 0.7 }_{- 0.6 }$ & 5.3 & 3.7 $\pm$ 1.0 & 12.1 $\pm$ 9.1 \\
SF\_M10.9 & 12.5 & 3.2 & 1.2 & 2.0 $^{+ 0.6 }_{- 0.5 }$ & 3.3 & 3.1 $\pm$ 1.0 & 10.1 $\pm$ 8.7 \\
\hline
SF\_M10.86 & 12.4 & 3.0 & 1.1 & 1.9 $^{+ 0.6 }_{- 0.5 }$ & 3.5 & 2.3 $\pm$ 0.8 & 8.5 $\pm$ 7.7 \\
SF\_M10.74 & 12.3 & 2.8 & 0.6 & 1.6 $^{+ 0.5 }_{- 0.4 }$ & 3.3 & 2.1 $\pm$ 0.7 & 4.3 $\pm$ 7.5 \\
SF\_M10.63 & 12.1 & 2.8 & 0.7 & 1.3 $^{+ 0.4 }_{- 0.3 }$ & 3.1 & 2.0 $\pm$ 0.7 & 5.3 $\pm$ 7.3 \\
SF\_M10.52 & 12.0 & 1.8 & 0.6 & 1.2 $^{+ 0.3 }_{- 0.3 }$ & 0.8 & 1.2 $\pm$ 0.7 & 4.2 $\pm$ 7.0 \\
SF\_M10.41 & 11.9 & 2.5 & 0.8 & 1.0 $^{+ 0.3 }_{- 0.2 }$ & 1.5 & 1.6 $\pm$ 0.7 & 5.8 $\pm$ 6.9 \\
SF\_M10.3 & 11.8 & 1.9 & 0.5 & 0.9 $^{+ 0.2 }_{- 0.2 }$ & 1.8 & 1.2 $\pm$ 0.6 & 3.4 $\pm$ 6.5 \\
SF\_M10.18 & 11.8 & 2.9 & -0.0 & 0.7 $^{+ 0.2 }_{- 0.1 }$ & 1.3 & 1.7 $\pm$ 0.6 & -0.2 $\pm$ 6.2 \\
SF\_M10.03 & 11.6 & 2.1 & 0.5 & 0.6 $^{+ 0.2 }_{- 0.1 }$ & 1.4 & 1.1 $\pm$ 0.5 & 2.8 $\pm$ 5.9 \\
SF\_M9.86 & 11.5 & -0.2 & -0.4 & 0.4 $^{+ 0.1 }_{- 0.1 }$ & 0.3 & -0.1 $\pm$ 0.4 & -2.3 $\pm$ 5.3 \\
SF\_M9.64 & 11.4 & 1.3 & 0.3 & 0.3 $^{+ 0.1 }_{- 0.1 }$ & 0.3 & 0.5 $\pm$ 0.4 & 1.5 $\pm$ 4.8 \\
\hline
\end{tabular}
\tablefoot{
Cylindrical projected luminosity in units of $10^{40}$ erg s$^{-1}$ measured within projected distances of 80 and 300 kpc for each sample with the M1 mask. 
In the column S/N we report the S/N\ of the measurement. We consider to have a "detection" when the S/N\ is larger than 3, a "hint of detection" if the S/N is between 1 and 3, and an "upper limit" when the S/N is smaller than 1.
These are compared with the XRB model prediction of \citet{Aird2017MNRAS.465.3390A}. 
"max PS" is the luminosity obtained when integrating the point-source profiles (gray lines in the figures), which constitutes an upper limit to the luminosity that can be attributed to point-source emission. 
$ \log_{10}(M_{vir}) $ is the estimated mean halo mass obtained using the stellar to halo mass relation from \citet{moster13}. 
Table \ref{tab:LX:80kpc:M3} reports the same quantities as obtained when applying the `M3' mask.}
\end{table*}

\begin{table*}
    \centering  
    \caption{\label{tab:LX:80kpc:M3}Cylindrical projected X-ray luminosity as in Table  \ref{tab:LX:80kpc:M1}, but using the M3 mask.}
    \begin{tabular}{c |c c c |c c c c c c c c c c}
\hline \hline
 sample &  $ \log_{10}(M_{vir}) $   & \multicolumn{2}{c|}{S/N} & \multicolumn{4}{c}{$L_{X}$ [$10^{40}$ erg s$^{-1}$]} \\
 name   & halo & 80 & 300 & XRB & max PS &  $R_p$<80\,{\rm kpc}& $R_p$<300\,{\rm kpc} \\
\hline

ALL\_M10.7 & 12.2 & 13.1 & 4.0 & 1.1 $^{+ 0.3 }_{- 0.2 }$ & 11.3 & 8.8 $\pm$ 0.7 & 29.2 $\pm$ 7.2 \\
\hline
& \multicolumn{4}{c}{Fixed stellar mass selection} \\
\hline
QU\_M10.7 & 12.2 & 8.3 & 4.6 & 0.6 $^{+ 0.2 }_{- 0.1 }$ & 4.1 & 4.9 $\pm$ 0.6 & 31.8 $\pm$ 6.9 \\
SF\_M10.7 & 12.2 & 14.9 & 3.5 & 1.4 $^{+ 0.4 }_{- 0.3 }$ & 16.0 & 11.2 $\pm$ 0.7 & 25.5 $\pm$ 7.4 \\
\hline
& \multicolumn{4}{c}{Quiescent galaxies} \\
\hline
QU\_M11.71 & 15.0 & 11.8 & 10.4 & 7.1 $^{+ 2.1 }_{- 1.6 }$ & 111.4 & 97.4 $\pm$ 8.3 & 459.0 $\pm$ 44.0 \\
QU\_M11.58 & 14.5 & 9.2 & 7.6 & 5.3 $^{+ 1.6 }_{- 1.2 }$ & 83.9 & 74.2 $\pm$ 8.1 & 319.2 $\pm$ 42.0 \\
QU\_M11.54 & 14.4 & 6.3 & 5.5 & 4.8 $^{+ 1.4 }_{- 1.1 }$ & 57.9 & 42.3 $\pm$ 6.7 & 199.5 $\pm$ 36.6 \\
\hline
QU\_M11.35 & 13.8 & 11.0 & 7.0 & 3.0 $^{+ 0.9 }_{- 0.7 }$ & 25.0 & 19.5 $\pm$ 1.8 & 89.8 $\pm$ 12.8 \\
\hline
QU\_M11.2 & 13.2 & 8.9 & 4.6 & 2.1 $^{+ 0.6 }_{- 0.5 }$ & 11.4 & 9.9 $\pm$ 1.1 & 44.6 $\pm$ 9.7 \\
\hline
QU\_M11.04 & 12.8 & 8.3 & 4.7 & 1.5 $^{+ 0.4 }_{- 0.3 }$ & 5.5 & 7.0 $\pm$ 0.8 & 40.3 $\pm$ 8.6 \\
QU\_M10.88 & 12.5 & 7.1 & 5.6 & 1.0 $^{+ 0.3 }_{- 0.2 }$ & 5.1 & 5.8 $\pm$ 0.8 & 47.2 $\pm$ 8.5 \\
QU\_M10.73 & 12.3 & 6.6 & 3.3 & 0.7 $^{+ 0.2 }_{- 0.1 }$ & 4.2 & 4.8 $\pm$ 0.7 & 25.0 $\pm$ 7.5 \\
QU\_M10.53 & 12.0 & 5.5 & 3.6 & 0.4 $^{+ 0.1 }_{- 0.1 }$ & 2.5 & 3.3 $\pm$ 0.6 & 23.7 $\pm$ 6.7 \\
\hline
& \multicolumn{4}{c}{star-forming galaxies} \\
\hline
SF\_M11.25 & 13.4 & 15.8 & 5.5 & 4.0 $^{+ 1.3 }_{- 1.0 }$ & 62.8 & 38.4 $\pm$ 2.4 & 78.0 $\pm$ 14.2 \\
SF\_M11.12 & 13.0 & 13.9 & 5.3 & 3.1 $^{+ 1.0 }_{- 0.7 }$ & 55.2 & 30.6 $\pm$ 2.2 & 73.5 $\pm$ 13.8 \\
SF\_M11.05 & 12.8 & 8.4 & 3.9 & 2.6 $^{+ 0.8 }_{- 0.6 }$ & 26.5 & 16.8 $\pm$ 2.0 & 51.1 $\pm$ 13.0 \\
\hline
SF\_M10.99 & 12.7 & 11.4 & 3.3 & 2.4 $^{+ 0.7 }_{- 0.6 }$ & 24.6 & 15.6 $\pm$ 1.4 & 33.4 $\pm$ 10.0 \\
SF\_M10.9 & 12.5 & 12.5 & 5.3 & 2.0 $^{+ 0.6 }_{- 0.5 }$ & 18.3 & 16.0 $\pm$ 1.3 & 52.3 $\pm$ 9.9 \\
\hline
SF\_M10.86 & 12.4 & 15.3 & 4.8 & 1.9 $^{+ 0.6 }_{- 0.5 }$ & 21.8 & 15.1 $\pm$ 1.0 & 41.6 $\pm$ 8.6 \\
SF\_M10.74 & 12.3 & 14.7 & 3.2 & 1.6 $^{+ 0.5 }_{- 0.4 }$ & 21.7 & 15.4 $\pm$ 1.0 & 26.0 $\pm$ 8.2 \\
SF\_M10.63 & 12.1 & 10.2 & 2.0 & 1.3 $^{+ 0.4 }_{- 0.3 }$ & 14.4 & 8.9 $\pm$ 0.9 & 15.3 $\pm$ 7.8 \\
SF\_M10.52 & 12.0 & 10.6 & 3.8 & 1.2 $^{+ 0.3 }_{- 0.3 }$ & 10.5 & 9.1 $\pm$ 0.9 & 30.2 $\pm$ 7.9 \\
SF\_M10.41 & 11.9 & 8.0 & 1.8 & 1.0 $^{+ 0.3 }_{- 0.2 }$ & 7.9 & 6.2 $\pm$ 0.8 & 13.3 $\pm$ 7.4 \\
SF\_M10.3 & 11.8 & 4.9 & 0.7 & 0.9 $^{+ 0.2 }_{- 0.2 }$ & 5.1 & 3.3 $\pm$ 0.7 & 4.8 $\pm$ 6.8 \\
SF\_M10.18 & 11.8 & 9.1 & 1.4 & 0.7 $^{+ 0.2 }_{- 0.1 }$ & 4.8 & 6.4 $\pm$ 0.7 & 9.5 $\pm$ 6.6 \\
SF\_M10.03 & 11.6 & 4.3 & 1.5 & 0.6 $^{+ 0.2 }_{- 0.1 }$ & 3.2 & 2.5 $\pm$ 0.6 & 9.5 $\pm$ 6.3 \\
SF\_M9.86 & 11.5 & 1.5 & 1.6 & 0.4 $^{+ 0.1 }_{- 0.1 }$ & 1.1 & 0.7 $\pm$ 0.5 & 9.4 $\pm$ 5.8 \\
SF\_M9.64 & 11.4 & 1.8 & 1.1 & 0.3 $^{+ 0.1 }_{- 0.1 }$ & 0.4 & 0.7 $\pm$ 0.4 & 5.7 $\pm$ 5.1 \\
\hline
\end{tabular}
\end{table*}
%
%
%
%
\section{Results}
\label{sec:result}

We discuss first the detection in the stacking experiment for the full sample (Sect. \ref{subsec:full:stack}). 
We consider to have a `detection' when the S/N is larger than 3, a `hint of detection' if the S/N is between 1 and 3, and an `upper limit' when the S/N is smaller than 1.

In Sections \ref{subsec:sfrBins:massBin} and \ref{subsec:trends}, we discuss the trends obtained when splitting the sample according to its sSFR and stellar mass. 
The comparison with theoretical predictions presented in Sect. \ref{sec:simulation} is done on the binned samples, where the stellar population is best controlled.

\subsection{Detection in the complete stack}
\label{subsec:full:stack}

We first report the results of our stacking exercise applied to the sample of 16\,142 galaxies at a mean redshift of 0.22 and a mean stellar mass of 10.7 (named ALL\_M10.7). We focus here on the results obtained by three possible masking procedures: ALL, M1, and M3 (see definitions in Sect. \ref{subsec:masking}).

When applying the ALL mask, we obtain a detection above the background; see top left panel of  Fig. \ref{fig:full:stack:measurement}. 
The S/N is $\sim$3 within $R_p<80$ kpc. 
At larger radii ($>80$ kpc), the signal measured (magenta crosses) is consistent with the background (green dashes). 
The shape of the profile is marginally more extended than the maximal point-source profile (gray step line).

The detection significance increases when using the M1 mask (Fig. \ref{fig:full:stack:measurement}, top middle panel), that is, when the galaxy clusters and groups detected by eROSITA in the redshift range of the GAMA galaxies  are not removed before stacking. 
The S/N accumulated within $R<80$ kpc is about $\sim$5 (reported in Table \ref{tab:LX:80kpc:M1}). 
Compared to the ALL mask profile, the M1 is brighter and significantly deviates from the maximal point-source profile. 

When using the M3 mask, that is, when the galaxy clusters, groups, {and} point sources detected by eROSITA among GAMA galaxies are not masked, the S/N within 80 kpc increases to $\sim$13; see Fig. \ref{fig:full:stack:measurement} (top right panel), and Table \ref{tab:LX:80kpc:M3}. 
The overall stacked profile corresponds, qualitatively, to what is expected with the addition of one (or multiple) bright unresolved source(s).

Finally, to measure the mean projected emission coming from around the galaxies, we subtract the background from each stacked profile; see the bottom panels of Fig. \ref{fig:full:stack:measurement}. 
There, background-subtracted profiles are shown out to 300 kpc. 
The possible deviation from a point-source emission profile is made clearer by the comparison with the corresponding "max PS" profile. 
Using the background-subtracted profiles, we measure the integrated projected luminosity in an aperture R$_p$ (in kpc) as follows: 
\begin{equation}
L^{<R_p}_X=\int_{r=0}^{r=R_p} [S_X(r)-BG]\; 2\pi \; r \; dr\,,
\end{equation}
where $BG$ is the background level estimated as in Sect. \ref{subsec:randoms}.
For each sample, the measured luminosity is reported for two apertures: 80 and 300 kpc in Tables \ref{tab:LX:80kpc:M1} (for mask M1) and  \ref{tab:LX:80kpc:M3} (for mask M3). 

The expected total (or radially integrated) XRB luminosity for the ALL\_M10.7 sample is 1.1$\times10^{40}$ erg s$^{-1}$. 
For the M1 (M3) mask, the luminosity measured in the inner 80 kpc is 2.8$\pm0.5 \times10^{40}$ (8.8$\pm0.7 \times10^{40}$) erg s$^{-1}$; therefore, the observed luminosity cannot come from XRBs alone. 

With the complete stack being constituted of a varied mix of different galaxies, linking the detection to underlying physical processes is complex. 
To further interpret the link between the detected emission and its possible sources (hot gas, XRBs, faint AGN), we split the GAMA sample according to the physical properties of the galaxies, as we describe in the following sections.

\begin{figure*}
    \centering
\includegraphics[width=.8\columnwidth ]{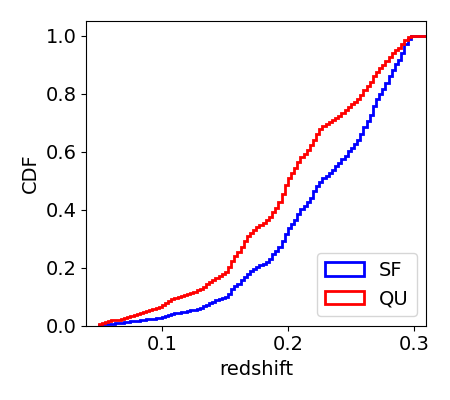}
\includegraphics[width=.8\columnwidth ]{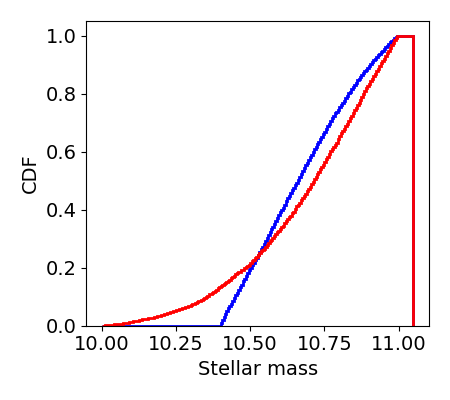}
    \caption{Redshift and stellar mass {cumulative normalized} distributions of the QU\_M10.7 and SF\_M10.7 samples. 
    Although the samples share a similar mean stellar mass (10.66 and 10.72) and a similar mean redshift (0.2 and 0.23), the underlying distributions differ.}
    \label{fig:subm1:107:comparison:hist:comparison}
\end{figure*}

\begin{figure*}
    \centering
\includegraphics[width=.9\columnwidth]{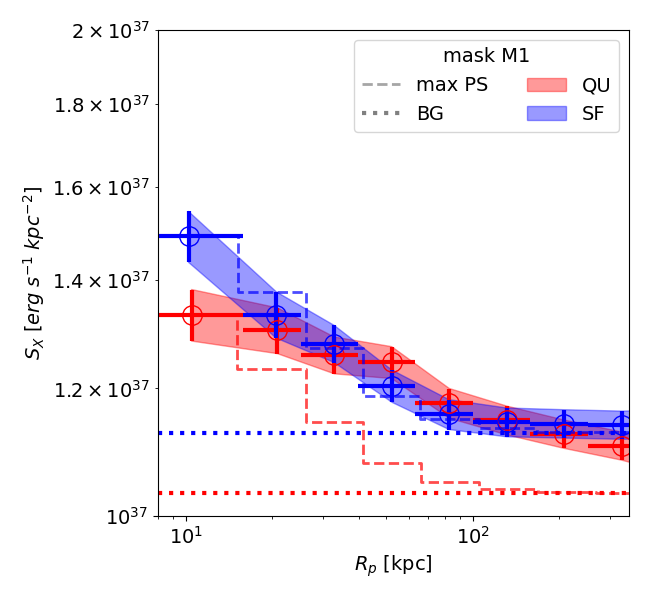}
\includegraphics[width=.9\columnwidth]{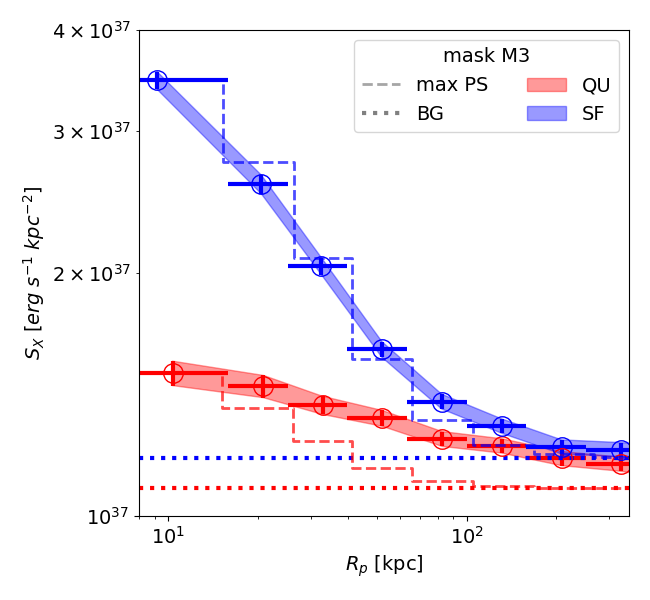}
\includegraphics[width=.9\columnwidth]{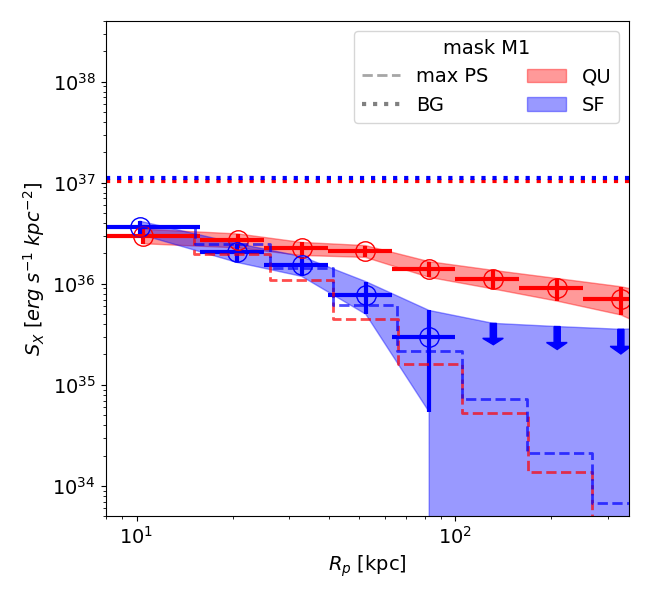}
\includegraphics[width=.9\columnwidth]{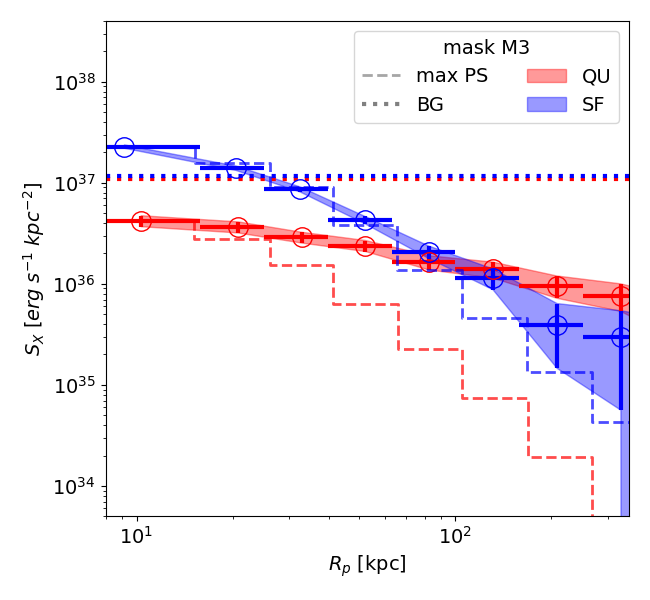}

    \caption{{Measured X-ray radial projected luminosity profiles (0.5--2.0 keV rest-frame) for the quiescent sample "QU\_M10.7" (red) and the star-forming sample "SF\_M10.7" (blue) of central galaxies, i.e., at similar median stellar mass around the scale of the Milky Way and Andromeda, albeit in the range $0.05<z<0.3$, using the M1 mask (which removes all X-ray-bright AGN in the samples; left column) and M3 mask (which does not remove X-ray-bright AGN from the sample; right column). Both samples have the same mean stellar mass of $\sim 10^{10.7} M_\odot$ and mean redshift of $z\sim0.2$ (but see the underlying distributions in Fig.~\ref{fig:subm1:107:comparison:hist:comparison}). QU\_M10.7 (M1 or M3) shows a clearly extended profile, while SF\_M10.7 (M1 or M3) shows a profile compatible with that of a point source convolved with the eROSITA PSF (dashes). The background level for each stack is given by the dotted lines. The bottom row shows the corresponding background-subtracted profiles.}}
    \label{fig:full:stack:SF:QU}
\end{figure*}

\subsection{Trend with sSFR at fixed stellar mass $\sim 5\times10^{10}$ M$_\odot$}
\label{subsec:sfrBins:massBin}

We split the sample into a quiescent subsample and a star-forming subsample. 
We set a maximum boundary in stellar mass at $10^{11}M_\odot$, which corresponds to haloes with a mass of $\sim5\times10^{12}M_\odot$ \citep[using the stellar-to-halo-mass relation from][]{moster13}, well below the halo mass of groups and clusters.  
We then search for the lower stellar mass boundary so that both samples have a mean mass of $\log_{10}(M^*/M_\odot)\sim10.7$, around the turn over (knee) of the stellar mass function, often called $\mathcal{M}^*$ \citep[][]{Ilbert2013AA...556A..55I}. 
We name these samples QU\_M10.7 for the quiescent galaxies and SF\_M10.7 for the star-forming galaxies. 
The mean redshift of each sample is close: 0.2 for QU\_M10.7 and 0.23 for SF\_M10.7, that is, a 0.3 Gyr difference. 
The sSFR of QU\_M10.7 is more than 100 times lower than that of  SF\_M10.7. Their redshift and stellar mass {normalized cumulative distributions} are shown in Fig. \ref{fig:subm1:107:comparison:hist:comparison}; {the SF sample is biased towards higher redshift and higher stellar mass} compared to the QU sample. 

The galaxies in these samples typically reside in dark-matter haloes of $5-50\times10^{11}\;M_\odot$ with a mean at $\sim16\times10^{11}\;M_\odot$, where the conversion of baryons into stars is thought to be the most efficient \citep{behroozi13a, moster13}. 
At this stellar mass, \citet{Velander2014MNRAS.437.2111V} found that the red galaxies reside in denser environments than blue galaxies. 
These latter authors measured the host halo mass of blue and red $\sim\mathcal{M}^*$ galaxies, and found that it was comprised between 1-3$\times10^{12}$M$_\odot$ for both. 
Based on these results, the samples QU\_M10.7 and SF\_10.7 are hosted on average by similarly massive haloes, and, importantly, reside in different environments. 

\subsubsection*{With the M1 mask}

In both star-forming and quiescent cases, there is an X-ray detection in the stacks obtained with the M1 mask; see Fig. \ref{fig:full:stack:SF:QU} {(left panels)}. 
For the QU\_M10.7 (SF\_M10.7) sample, the cumulative S/N within 80 and 300 kpc is of 7.3 (3.6) and 4.3 (0.9).
In the case of the QU\_M10.7 sample, the emission is clearly extended and not centrally peaked; see the red curves in the left panels of Fig. \ref{fig:full:stack:SF:QU}.
For the SF\_M10.7 sample, the emission is centrally peaked and consistent with a maximal PSF profile; see the blue curves in the left panels of Fig. \ref{fig:full:stack:SF:QU}. 
With a direct comparison of the background-subtracted profiles in Fig. \ref{fig:full:stack:SF:QU} ({bottom} left panel), the difference between QU\_M10.7 and SF\_M10.7 is made obvious. 
At scales larger than 100 kpc, the quiescent galaxy profile is at least two times brighter than that of the star-forming galaxies.

Our measurement, using a nearly complete galaxy catalog, provides firm observational evidence: at the same mean stellar mass of $\sim5\times 10^{10}M_\odot$, star-forming galaxies show significantly less projected X-ray emission on large scales in the 0.5--2 keV rest-frame energy range. The possibility of such a difference between passive and star-forming $\sim\mathcal{M}^*$ galaxies was previously suggested by \citet{bregman2018}, hinting at a difference in their evolutionary histories. 

The shallow slope measured in the QU profile might come from projection effects (see discussion in Sect. \ref{subsec:disc:projection}) due to the fact that quiescent galaxies tend to live in denser, hotter environments,
as mentioned above \citep[e.g.,][]{Velander2014MNRAS.437.2111V}.

\subsubsection*{With the M3 mask}

The right panels of Figure \ref{fig:full:stack:SF:QU} show the projected luminosity profiles obtained with the M3 mask. 
When including the detected X-ray point sources, the central parts of the profiles increase significantly for the SF profile, mainly because of the contribution of bright AGN (see Sect. \ref{subsec:agn}), while for the QU profile the increase is less noticeable. 
As opposed to the M1 stack, there is a hint of extended emission in the SF background-subtracted profile; {see the bottom
right panel of Fig. \ref{fig:full:stack:SF:QU}}. 

\subsubsection*{Projected luminosity within 80 and 300 kpc}

The integrated projected luminosity (within 80 and 300 kpc apertures) is reported in Tables \ref{tab:LX:80kpc:M1} (M1) and \ref{tab:LX:80kpc:M3} (M3). With the M1 mask, for QU\_M10.7, the emission within 80 kpc is 
$L_X = 4.0 \pm 0.5 \times 10^{40}$ erg s$^{-1}$, 
almost an order of magnitude greater than the corresponding prediction for the unresolved XRB luminosity from \citet[][]{Aird2017MNRAS.465.3390A} of $L_X = 6 \times 10^{39}$ erg s$^{-1}$. 
Given this, and the fact that the  profile of the QU\_M10.7 galaxies is clearly extended, we may conclude that this emission is mainly coming from both the hot gas in the CGM of the individual galaxies and from the projection of the large-scale environment around them. 
Concurrently, the SF\_M10.7 emission within 80 kpc amounts to $ 1.9 \pm 0.5 \times 10^{40}$ erg s$^{-1}$, which is compatible within 1$\sigma$ with that expected from XRBs ($1.4^{+0.4}_{-0.3} \times 10^{40}$ erg s$^{-1}$).
The profile also does not appear extended.
We are led to conclude that around star-forming galaxies of a mean mass of $\log_{10}(M^*/M_\odot)\sim10.7$, an extended hot gas component is not significantly detected. We refer the reader to Sect. \ref{subsubsec:agn_cont_discussion} for a more comprehensive discussion of the AGN contamination.

{With the M3 mask,  where the stack includes all X-ray-bright detected AGN, the luminosity of the QU sample increases by 20\% compared to the M1 mask, while for the SF sample, it increases by a factor of almost 6.} 
There may be a hint of extended emission in the SF profile measured on the larger scales between 200 and 300 kpc. 

\begin{figure*}[h!]
    \centering
\includegraphics[width=.9\columnwidth]{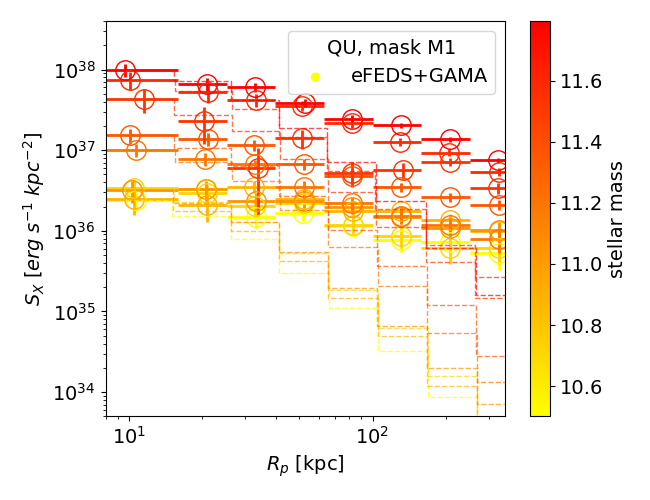}
\includegraphics[width=.9\columnwidth]{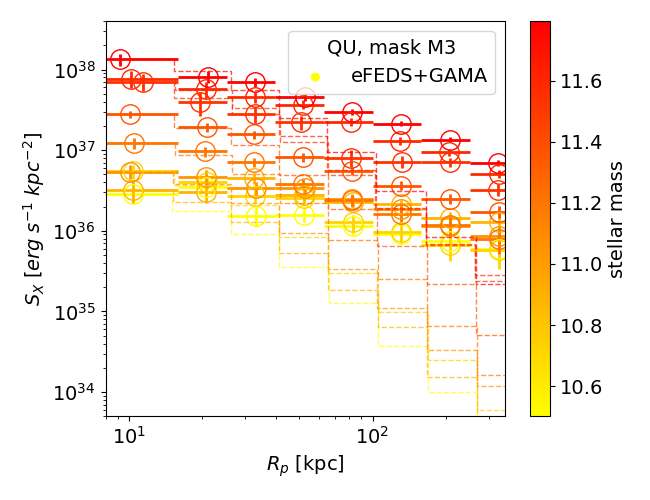}
\includegraphics[width=.9\columnwidth]{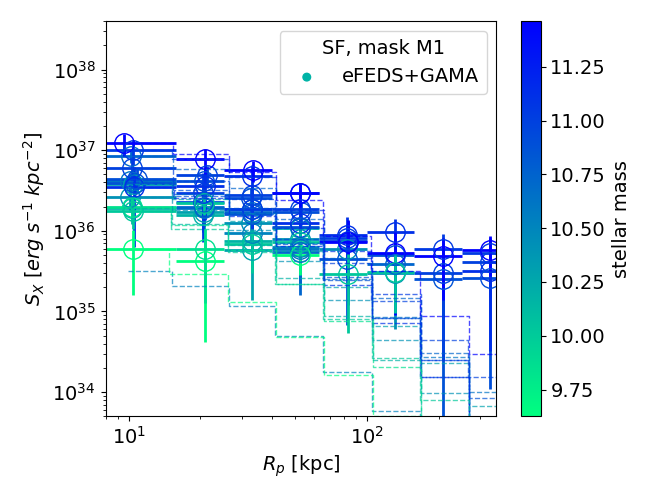}
\includegraphics[width=.9\columnwidth]{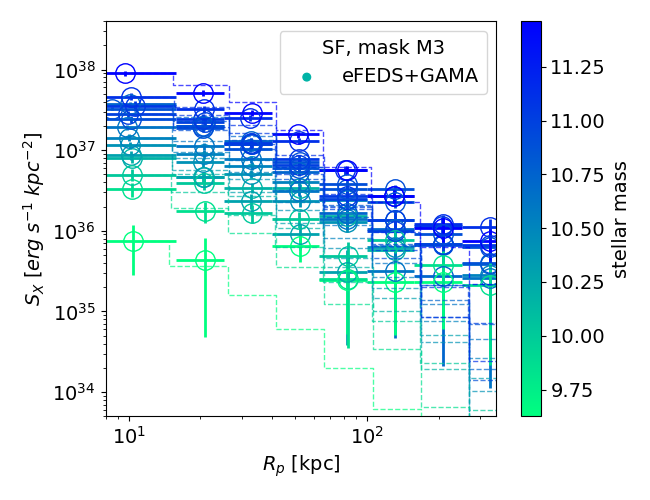}
    \caption{Comparison of the background-subtracted projected luminosity profiles in the 0.5--2.0 keV rest-frame band (M1 mask, left panels; M3 mask, right panels) for quiescent (top) and star-forming (bottom) samples as a function of galaxy stellar mass. 
    {The star-forming profiles are compatible with point-source emission profiles (dashes), although we note that the uncertainties are large, in particular for projected radii larger than 100 kpc. The quiescent profiles appear extended in comparison}.}
    \label{fig:subm1:trends}
\end{figure*}

\subsection{Trends with stellar mass}
\label{subsec:trends}

We further investigate trends as a function of stellar mass and sSFR with the set of samples specified in Table \ref{tab:gama:data}. 
For the quiescent samples, we are limited by the total number of galaxies available in the catalog and we are not able to create lower mass bins. 
For the star-forming samples, we define samples down to stellar masses of $3\times 10^9$ M$_\odot$.
For the M1 mask, we report a detection (S/N$>3$ in 80 or 300 kpc) for all quiescent samples and only for a handful of the star-forming galaxy samples; see Table \ref{tab:LX:80kpc:M1}. For the M3 mask, we report a detection for all samples except for the star-forming samples with a stellar mass lower than $10^{10} M_\odot$; see Table \ref{tab:LX:80kpc:M3}. 

The sets of background-subtracted projected luminosity profiles obtained in the M1  and M3 masks are shown in the left and right panels of Fig. \ref{fig:subm1:trends}, respectively. 
The qualitative trend observed for the quiescent samples (M1 or M3 mask) is in line with expectations: the higher the stellar mass (and therefore the host halo mass), the brighter the emission and the higher the S/N (Tables \ref{tab:LX:80kpc:M1}, \ref{tab:LX:80kpc:M3}). 
For the star-forming samples, with the M1 mask, all profiles except those at the highest mass end are broadly consistent with one another on large scales and are dominated by noise. 
A difference between profiles arises in the first radial bins; for example, the mean brightness of the central point source scales with the stellar mass. 
For the star-forming samples with the M3 mask, the amplitude correlates with stellar mass at all scales, as in the QU profiles.
This follows the expectation that the mean AGN luminosity is correlated with the host stellar mass \citep{Aird2017MNRAS.465.3390A, Comparat2019MNRAS.487.2005C, Georgakakis2019MNRAS.487..275G}. 
Possible extended emission around star-forming galaxies remains to be significantly detected. 

\subsection{Scaling between X-ray projected luminosity and stellar mass:  M1 mask}

\begin{figure*}[h!]
    \centering
\includegraphics[width=1.6\columnwidth]{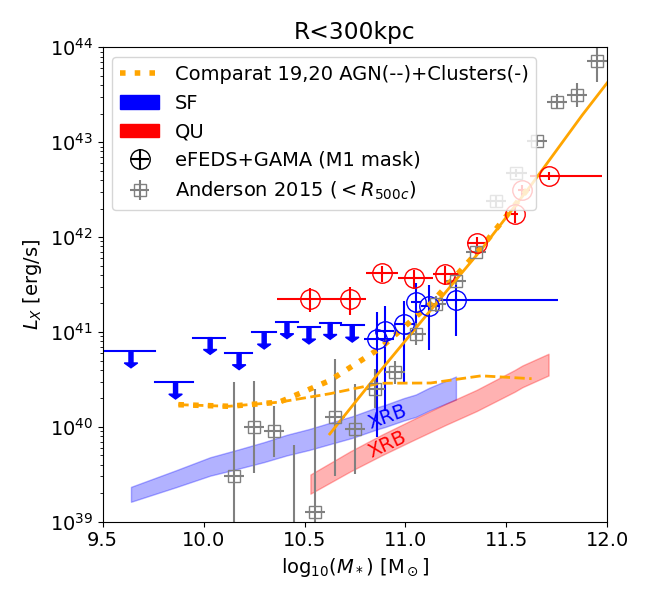}
\includegraphics[width=.95\columnwidth]{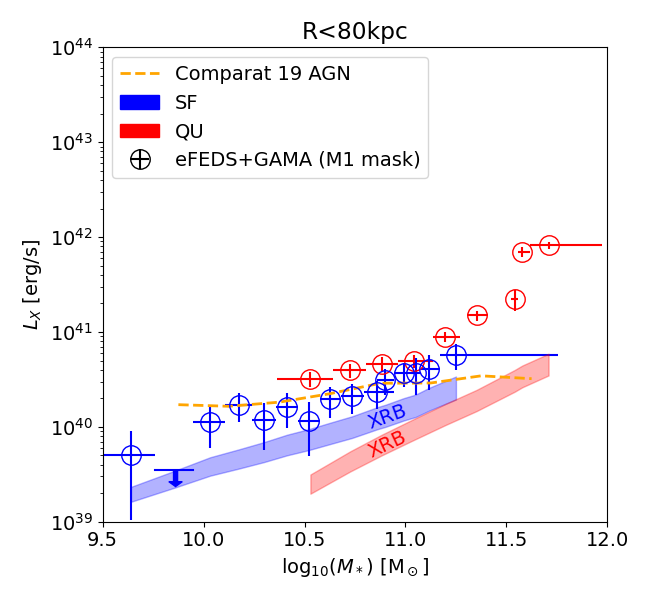}
\includegraphics[width=.95\columnwidth]{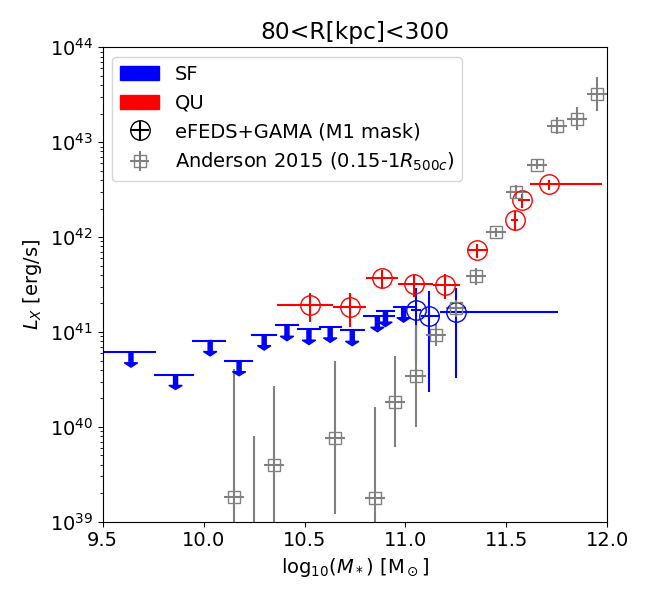}

    \caption{X-ray 0.5--2 keV projected luminosity around central galaxies as a function of galaxy stellar mass and split into star-forming (blue symbols and annotations) and quiescent (red) samples, computed  using the  M1 mask. Each eFEDS+GAMA detection is indicated with circles, upper limits with downwards arrows. In the \textit{Main panel}, we show the luminosity integrated within 300 physical projected kpc.
    In the \textit{Bottom Left Panel}, we show the luminosity integrated within 80 physical projected kpc. 
    In the \textit{Bottom Right Panel}, the relation is shown for the outer shell 80-300 projected kpc. Gray squares are the measurements from \citet{anderson15}, computed within $R_{500c}$ (main panel) and within 0.15--1 $R_{500c}$ (bottom right panel). The orange dashed line shows the prediction from the AGN population synthesis model (after excluding sources with $F_X>6.5\times 10^{-15}$ erg s$^{-1}$ cm$^{-2}$, as per M1 mask) of \citet{Comparat2019MNRAS.487.2005C}. {The orange solid line shows the prediction for the clusters and groups using the model of \citet{Comparat2020OJAp....3E..13C}, i.e., the contribution of hot virialized haloes. The dotted line is the sum of the two.}}
    \label{fig:scaling:relation}
\end{figure*}

Figure \ref{fig:scaling:relation} shows the scaling measured between X-ray luminosity and stellar mass within 300 kpc (main panel), in the inner 80 kpc (bottom left), and in the shell 80-300 kpc (bottom right), all obtained with the M1 mask applied. 
Overall, the S/N is highest in the central 80kpc; see Table \ref{tab:LX:80kpc:M1}. 
It decreases when integrating to 300kpc. 
Indeed extending the integration to larger scales, the signal increases marginally while the noise increases much more, resulting in lower S/N.
We find that X-ray luminosity correlates with mean stellar mass. The trend for star-forming galaxies is different from that of quiescent galaxies. 
However, there appear to be two regimes in the scaling between X-ray projected luminosity and stellar mass. The emission from the inner parts  is dominated by point sources (AGN and XRB), while that from the outer parts is dominated by CGM emission. 

In particular, within 80 kpc, the slope of both SF and QU galaxies is similar to (but offset from) that predicted for XRBs, and consistent with the predicted unresolved AGN population \citep[orange shaded area,][]{Comparat2019MNRAS.487.2005C}. 
The AGN population is predicted using eROSITA mock catalogs filtered on X-ray flux and optical magnitude to exclude the X-ray AGN that are optically brighter than the magnitude limit of GAMA: $F_X<6.5\times10^{-15}$ erg cm$^{-2}$ s$^{-1}$ and $r<19.8$. 
Those simulated AGN could be hosted by GAMA galaxies but would not have been detected in eFEDS. 
The simulations used start to be incomplete at stellar masses of $10^{10}M_\odot$ at $z=0.22$, and we therefore limit the prediction to above this mass.

Still within 80 kpc (bottom left panel), only for the highest stellar mass quiescent galaxy samples, that is, for stellar mass $>2\times10^{11}$, corresponding to a halo mass  $\gtrsim5\times10^{13}$, do we measure a luminosity that is significantly brighter than the predicted point-source emission. This is due to the large amount of hot gas in projection present in galaxy groups and galaxy clusters.

Within 300 kpc, the X-ray luminosity measured around SF samples is consistent with the predicted average point-source emission (combination of AGN plus XRBs, dominated by AGN emission). 
For stellar masses above $10^{11}$, the emission is marginally brighter than the expected point-source contribution. 

We now consider the results for the 80-300 kpc shell shown in the bottom right hand panel of Fig. \ref{fig:scaling:relation}. 
For the quiescent sample and stellar masses above $\log M^*\sim11.2$, the measurements are in good agreement with \citet{anderson15}\footnote{The slight discrepancy at the highest mass is likely due to the difference in aperture: {500c} is larger than 300 kpc for a 10$^{15}M_\odot$ halo.}. 
Below $\log M^*\sim11.2$, the luminosity measured is significantly above that of \citet{anderson15}.
We believe this is due to projection effects for the QU sample, which preferentially resides in dense and hot environment. We discuss this effect in Sect. \ref{subsec:disc:projection}.
Still in the 80-300 kpc shell, for the star-forming samples, we only measure upper limits to the extended emission, except for the three highest stellar mass samples, where, on the other hand, the error bars extend to a low-luminosity value, meaning only marginal detection, with S/N $\sim$1.3.

\subsection{Scaling between X-ray projected luminosity and stellar mass: M3 mask}

\begin{figure*}[!]
    \centering
\includegraphics[width=1.6\columnwidth]{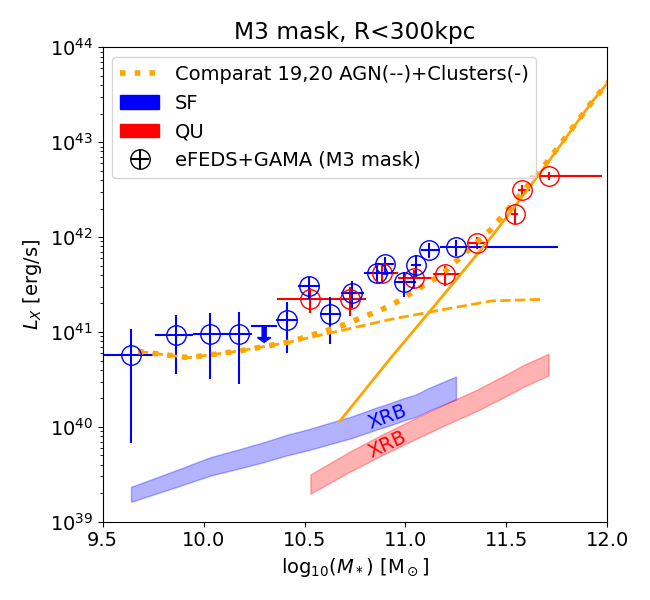}

    \caption{Same as Fig. \ref{fig:scaling:relation} but for the M3 mask (i.e., including detected point sources). 
    Predictions based on the empirical AGN and cluster models from \citet{Comparat2019MNRAS.487.2005C, Comparat2020OJAp....3E..13C} (now including sources with $F_X>6.5\times 10^{-15}$ erg s$^{-1}$ cm$^{-2}$, as per M3 mask) {are shown as an orange dashed line (AGN) and  a solid orange line (groups and clusters) and its sum (thick dotted orange line)}. The agreement between model and observations is remarkable. For stellar masses between $2\times10^{10}M_\odot$ and $2\times10^{11}M_\odot$, the positive offset between the observations and models is likely related to a combination of emission from the CGM and projection effects.}
    \label{fig:scaling:relation:EMP:model}
\end{figure*}

The relation obtained with the M3 mask is to be interpreted as the sum of all emitting entities: AGN, XRBs, and hot gas augmented by systematic projection effects. In that regard, there is no need to split as a function of projected separation. 
Figure \ref{fig:scaling:relation:EMP:model} shows the scaling measured in the inner 300 kpc with the M3 mask applied. 

We predict the AGN, the galaxy group, and the galaxy cluster population using the eROSITA mock catalog methods \citep{Comparat2019MNRAS.487.2005C,Comparat2020OJAp....3E..13C,Liu_SIM_2021arXiv210614528L,Seppi2022arXiv220709242S}. 
For the AGN population, we select all X-ray AGN that are optically brighter than the magnitude limit of GAMA: $r<19.8$. 
These model AGN could be hosted by GAMA galaxies regardless of whether  or not they are detected in eFEDS. 
No filter is applied for the cluster and group population. 
The black dashes represent the sum of the AGN and cluster contribution to this relation. 
{The sum of the two empirical models should correspond to the relation measured when applying the M3 mask, that is, when all detected sources are left in the stack.} 
We find that the luminosity--stellar mass relation is in good agreement with the models, which demonstrates that the models of \citet{Comparat2019MNRAS.487.2005C, Comparat2020OJAp....3E..13C}, \citet{Seppi2022arXiv220709242S} accurately represent the large-scale structure seen in X-rays. 

At high mass, above $2\times10^{11}M_\odot$, the measurements are slightly below the model. This is likely due to the fixed 300 kpc aperture used, which for these masses is smaller than the $R_{500c}$ used in the cluster model. 
For masses below $2\times10^{10}M_\odot$, measurements are consistent with the AGN model prediction, meaning that a detection of CGM emission is unlikely.
For stellar masses between $2\times10^{10}M_\odot$ and $2\times10^{11}M_\odot$, the positive offset between the observations and models is likely related to emission from the CGM and to projection effects. 
Given the uncertainties on the measurement and the large scatter in the model prediction, the quantitative assessment of the difference between the observation and the models is a complex undertaking.  

\begin{figure*}
    \centering
\includegraphics[width=.9\columnwidth ]{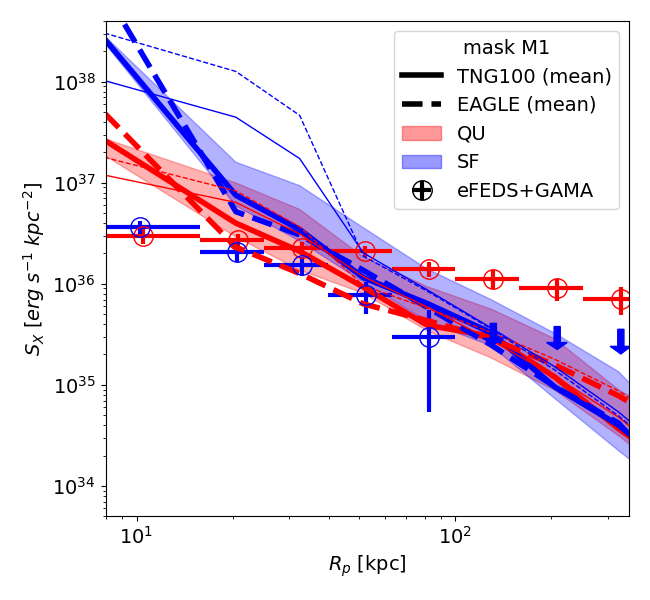}
\includegraphics[width=.9\columnwidth ]{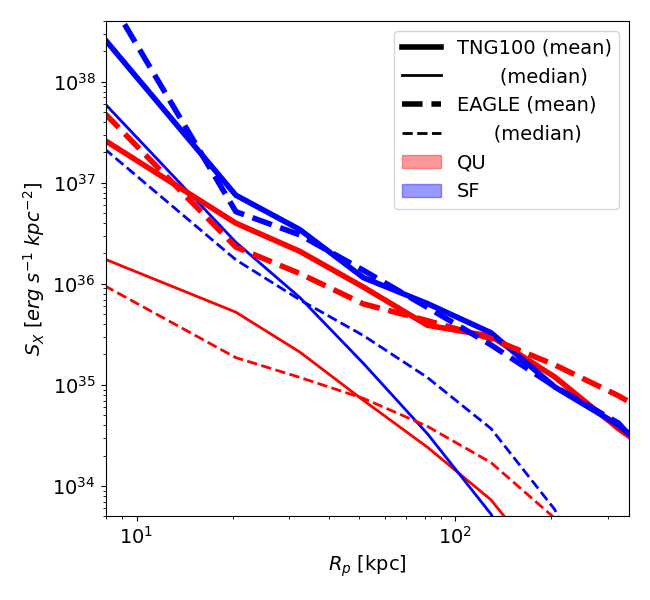}
    \caption{{\it Left:} X-ray luminosity projected radial profiles (0.5--2.0 keV rest-frame) for the quiescent QU\_M10.7 and the star-forming SF\_M10.7 samples (as in Fig.~\ref{fig:full:stack:SF:QU} left panel), i.e., for central galaxies with median galaxy stellar mass of about $5\times10^{10}\,\MSUN$ and median redshift 0.20. Their mass and redshift distributions are shown in Fig.~\ref{fig:subm1:107:comparison:hist:comparison}. 
    The profiles are compared to the results from the TNG100 (solid) and the EAGLE (dashed) simulations, consistently matched in stellar mass, sSFR, and redshift. 
    Observed stacking results are compared to the mean simulated profiles (thick curves). 
    For illustrative purposes, we convolve the mean predicted simulated profiles (thick curves) with the PSF (dashed lines in Fig. \ref{fig:full:stack:SF:QU}) and obtain the thin curves (ideally this should be done on each individual simulated galaxy profile before taking the mean). 
    This provides a rough idea of how the flux in the inner regions is shifted to larger radii due to the PSF. 
    {\it Right:} Complete set of profiles predicted by the simulation: mean (thick  {lines}) and median (thin lines). The median profiles show where the majority of the simulated profiles are located; {the median curve is significantly lower than the mean curve}. Shaded areas represent the systematic uncertainties associated to the extraction of the mocked observable from the simulations (specifically for TNG100); see text for details. {We expect the possible systematic uncertainties to be similar for the EAGLE simulation, but we refrain from adding shaded areas in order to avoid overcrowding the figure}.
    }
    \label{fig:full:stack:SF:QU:SIMS}
\end{figure*}

\begin{figure*}[h!]
    \centering
    \includegraphics[width=1.6\columnwidth]{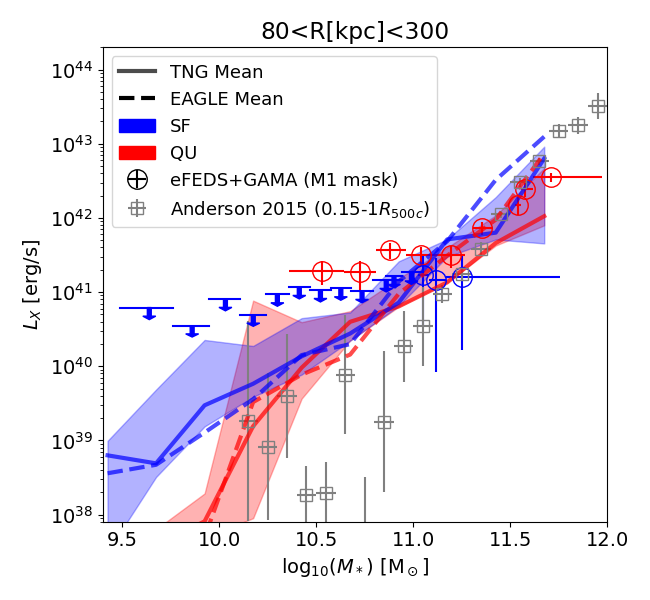}
    \includegraphics[width=.85\columnwidth]{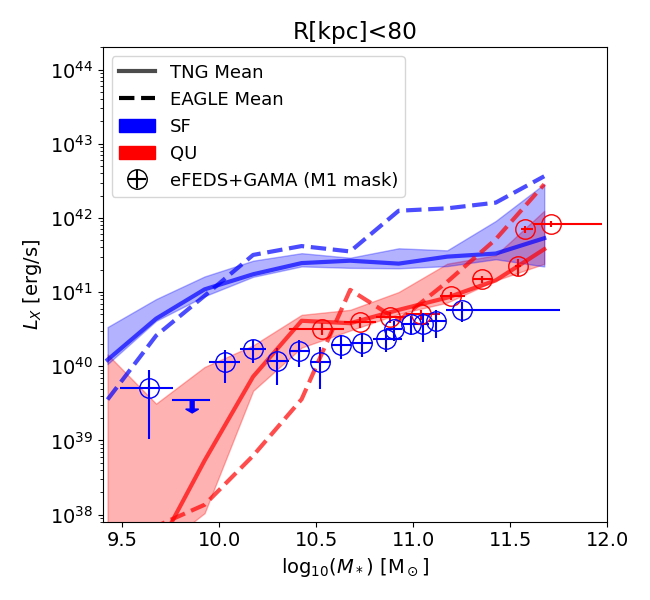}
    \includegraphics[width=.85\columnwidth]{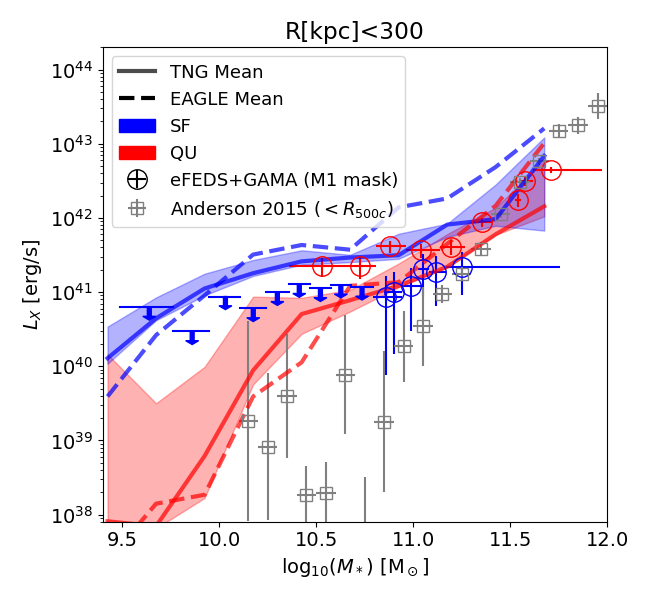}
    \caption{Same as Fig.~\ref{fig:scaling:relation} but with the addition of the predictions from the TNG100 and EAGLE simulations of matched galaxies. Shaded regions give the order of magnitude of the systematic uncertainty due to the process used to create mock observations. Shaded regions are shown for the TNG simulation. For EAGLE, shaded regions should have a similar width, but they are not shown so as to not overcrowd the figure. We show the observed and predicted soft X-ray luminosity from the central 80kpc (bottom left) from the full 300kpc (bottom right) and for the $80<R_p<300$ kpc range (main top panel) as a function of galaxy stellar mass.}
    \label{fig:scaling:relation:wSim}
\end{figure*}

%
%
%
%
\section{Comparison with simulated galaxies}
\label{sec:simulation}

We elect the IllustrisTNG \citep[hereafter TNG,][]{pillepich2018b,nelson2018,naiman2018,marinacci2018, springel2018} and the EAGLE simulations \citep{schaye2015,crain2015} as our reference points for the comparison of the results uncovered by eROSITA with the predictions from current state-of-the-art cosmological hydrodynamical simulations of galaxies. The reasons are manifold. 
First, both numerical projects provide flagship runs that encompass sufficiently large volumes and therefore sufficiently large numbers of galaxies for the construction of samples comparable to the ones inspected in this paper — there are 6\,478 and 3\,557 galaxies with galaxy mass $\log_{10} M^*>10$ in the TNG100 (TNG) and Ref-L0100N1504 (EAGLE) boxes, respectively, all at $z=0$. This would not be the case with zoom-in projects, which for massive galaxies are limited to examples of a few to a few tens of sources. 
Second, their outcomes have been contrasted to an ever-increasing set of observables, with galaxy populations at low and intermediate redshift that are well within ($<$1 dex) the range of  the observational constraints both in terms of demographics and inner galaxy properties. For example, the TNG simulations have been shown to accurately reproduce the low-redshift results obtained from Sloan Digital Sky Survey data in relation to: the $(g-r)$ color distributions across galaxy masses \citep{nelson2018}; the quiescent fractions of both centrals and satellites as a function of stellar mass \citep{Donnari2021}; and the small- and large-scale spatial clustering of galaxies, also when split by galaxy color \citep{springel2018}.
Third, in both cases, predictions for the X-ray emission of the gas within and around galaxies at $z\sim0$ have already been extensively quantified across a wide range of masses, galactocentric distances, and for star-forming and quiescent galaxies separately \citep{Truong20, Oppenheimer2020}. 
For example, in the 0.5--2 keV band, the $L_{\rm X}( < \rm{R}_{\rm 500c})$ versus M$_{\rm 500c}$ scaling relations of TNG are within the observational constraints provided by for example \citet{Pratt2009AA, Vikhlinin2009ApJ...692.1033V}, \citet{Sun2012NJPh...14d5004S}, \citet{Mehrtens2012MNRAS.423.1024M}, \citet{Lovisari2015AA...573A.118L} throughout the $10^{13-15}\, \MSUN$ range \textcolor{cyan}{(Pop et al. in prep.)}. As are those of EAGLE \citep{barnes17}. Finally, TNG and EAGLE are publicly available \citep{nelson2019a, mcalpine2016}.

{
Haloes in the simulations are identified with the {Friends-of-Friends (FoF)} algorithm both in the case of TNG and EAGLE: no a priori cuts are placed to their extent. Galaxies within haloes are identified by searching for gravitationally bound structures within the FoF haloes with the SUBFIND algorithm. All this is described in detail in the aforementioned release papers.}

\subsection{Extraction of the CGM observables from the simulated data}

First, we construct simulated galaxy samples that are matched to the observed ones by finding, for each galaxy in the GAMA set, its simulated equivalent in TNG and EAGLE. The details of this procedure are given in Appendix A. In practice, results are shown by averaging across 20 Monte Carlo samples of TNG100 and EAGLE galaxies matched to the GAMA sample adopted in this paper.

Second, X-ray photons are not explicitly modeled by the TNG and EAGLE simulations. However, the X-ray intrinsic luminosity that would be emitted by the simulated galaxies can be derived given the physical properties of the gas (i.e., of the plasma) returned and predicted by the numerical model. In practice, here we rely on the mapping between observed eROSITA photon count rates and X-ray fluxes adopted throughout and described in Sect. \ref{sec:method}. However, we only model the X-ray emission from the volume-filling gas, that is, we do not attempt to model the contamination from XRBs or AGN.

For any gas cell or gas particle in the simulations, barring the star-forming ones and with each one being characterized by a density, temperature, and metallicity, we obtain the [0.5--2] keV luminosity assuming a single-temperature Astrophysical Plasma Emission Code, \texttt{APEC 3.0.9}, as implemented in the \texttt{XSPEC}\footnote{\texttt{https://heasarc.gsfc.nasa.gov/xanadu/xspec/}} \citep{smith01} package. There, we assume an optically thin plasma in collisional ionization equilibrium. For element abundances, we employ the table provided by \citet{Anders1989GeCoA..53..197A} 
 re-scaled by the overall metallicity of the gas cells\footnote{We checked that, by using the individual abundances of nine elements tracked by TNG100 instead of the overall metallicity, the X-ray profiles of $5\times 10^{10}$ M$_\odot$ galaxies vary by about 0.1 dex and overall by negligible amounts in comparison to other systematic uncertainties (as described in Sect.~\ref{Result_TNG_EAGLE}).}. 
 
{For  each  TNG100  and  EAGLE  galaxy  matched  to  the GAMA sample, we consider all the gas around it and that belongs to their FoF host halo, with no a priori cut to the spatial extent of the  gas. More specifically, we sum up the contribution to the total X-ray luminosity along the line of sight in a given projection by accounting for all the gas cells or particles within the FoF selection: such a line-of-sight projection can span between many hundreds of kpc to several Mpc depending on the halo mass.} To obtain projected X-ray profiles, we take the minimum of the potential as the galaxy center and we determine, for each individual galaxy, the X-ray luminosity as a function of radius in a random projection, that is, in a random galaxy orientation.
We mimic the stacking signal of a specific galaxy subsample by taking the average (mean) of the radially binned X-ray luminosity values from all the simulated galaxies in the matched subsample. As the eFEDS stacking profiles are de facto weighted by the photon counts in each radial bin, we believe that the {mean} profiles across individual simulated ones is the closest approximation to observed stacked signals. 

We convolve the mean simulation profiles with the eROSITA PSF, but we do not remove ---from the simulation data--- those unresolved sources that are indeed detected but then masked (M1) in the eFEDS+GAMA results\footnote{The latter task requires modeling the X-ray emission not only from the gaseous component but also from AGN and X-ray binaries, which is beyond the scope of the current paper.}.
Moreover, we defer the task of simulating eROSITA photons, as for example done for the tailored predictions of \citet{Oppenheimer2020},  to a future dedicated paper. With such a full forward modeling into the observational space, we will then also be in the position of replicating, with the simulation data,  the exact stacking procedure adopted here for the eFEDS+GAMA data.
This would help in particular to quantify the extent of projection effects along the line of sight. 

\subsection{Results from IllustrisTNG and EAGLE}
\label{Result_TNG_EAGLE}

Results for the TNG100 and EAGLE galaxies in comparison to eFEDS+GAMA inferences are shown in Figs.~\ref{fig:full:stack:SF:QU:SIMS} and \ref{fig:scaling:relation:wSim}: there we focus, respectively, on the projected radial profiles at the transitional mass regime of $5\times10^{10}\,\MSUN$ and on the integrated signal from the CGM as a function of galaxy stellar mass, that is, integrating the X-ray luminosity {within various apertures: $[0-80]$, $[0-300]$, and $[80-300]$ projected kpc, with more emphasis on the latter, that is, beyond the typical extent of the eROSITA PSF at the considered redshifts.} 
In both figures, shaded areas around TNG100 results quantify the systematic uncertainties in the sample-matching procedure. {Systematic uncertainties on the EAGLE simulation are expected to be similar.} These are obtained by encompassing the 5th--95th percentile results when: \textit{(i)} marginalizing over the 20 Monte-Carlo sampling realizations of the GAMA samples; \textit{(ii)} using the total galaxy stellar masses from the simulations versus those within smaller apertures: twice half stellar mass radius; and \textit{(iii)} using the instantaneous and inner SFR values 
from the simulations versus those averaged over the last 100 Myr. In the profiles of Fig.~\ref{fig:full:stack:SF:QU:SIMS}, for example, these systematic choices can amount to X-ray luminosity uncertainties of about 0.5--0.7 dex at 200--300 kpc projected radii.

\subsubsection*{Comparison between 80 and 300 projected kpc}

Focusing on the CGM, at galactocentric distances $\gtrsim80$ kpc {(beyond the eROSITA PSF)}, the mean X-ray profiles of MW- and M31-mass galaxies predicted by TNG100 (solid) and EAGLE (dashed curves) are very similar to one another, despite the different underlying galaxy physics models: they both fall within approximately 1 dex from the observational results. Moreover, the profiles of the simulated star-forming versus quiescent galaxies are not significantly different from one another in the simulations, with the simulated X-ray atmospheres around quiescent $10^{10.7}\,\MSUN$ galaxies being less luminous than what the observations imply in Fig.~\ref{fig:full:stack:SF:QU:SIMS}.

{The top panel in Fig.} \ref{fig:scaling:relation:wSim} shows a comparison between the observed extended (between 80 and 300 projected kpc) X-ray luminosity as a function of stellar mass and the simulation predictions for the sample  matched in redshift,  mass, and sSFR to the GAMA sample. As for the case of the radial profiles, the CGM luminosity in the soft-X-ray band as a function of galaxy stellar mass is not too dissimilar between TNG100 and EAGLE, with similar emission for quiescent and star-forming galaxies at fixed stellar mass in both models. 

For the quiescent massive galaxies, TNG100 and EAGLE are in good agreement with the observations, especially at $\gtrsim 2\times 10^{11}\MSUN$. 
{Star-forming galaxies at $\gtrsim 2\times 10^{11}\MSUN$ have luminosities below the TNG and EAGLE predictions: when considering the systematic uncertainties due to the matching procedure (0.5-0.7 dex) and the uncertainties of the observations, the upper limit of the measurement is at the limit of being consistent with the lower (simulation) limit.} 
For $\lesssim10^{11}\MSUN$ galaxies, simulations predict a lower luminosity than observed, which is consistent with the nondetection of extended emission around star-forming galaxies, but is significantly lower than the observed luminosity of the quiescent galaxies. 

We note that the simulations predict luminosities that are possibly greater than that observed by \citet{anderson15}. 
The difference between the simulated curves and the observations gives a sense of the maximal amount of luminosity that could be imputed to projection effects for quiescent galaxies.

\subsubsection*{Comparison below 80 projected kpc}

{It is manifest from Fig.~\ref{fig:full:stack:SF:QU:SIMS} that both TNG100 and EAGLE predict much brighter atmospheres at small galactocentric distances than what is found with eFEDS+GAMA: up to two orders of magnitude brighter profiles at $< 20$ kpc. This discrepancy is also seen at other mass scales, as shown in the bottom panels of Fig. \ref{fig:scaling:relation:wSim}, but to different degrees for star-forming and quiescent galaxies: the integrated X-ray emission within 80 kpc (as well as within 300 kpc) of simulated quiescent galaxies appears to be largely consistent with the observed values; on the other hand, the simulated star-forming galaxies exhibit gaseous haloes that are  systematically
brighter in  X-rays than the observed counterparts across the considered mass range ($M_*=10^{9.5-11.5}M_\odot$). \\
The simulations may over-predict the amount of hot gas in the central regions of the  galaxies, in particular for star-forming galaxies. In fact, the latter possibility is partially at odds with the conclusions of \citet[][, their Figure 6]{Truong20} where the X-ray luminosity within the stellar effective radii of TNG star-forming galaxies appears compatible with analogous measurements of individual star-forming galaxies in the local Universe by \citet{Mineo2014MNRAS.437.1698M}, although less so with data from \citet{li13}. It should be noted that the simulation signals only come from the volume-filling gas (with no contributions from the hot ISM), whereas in the observations, part (if not all) of the signal comes from unresolved point sources. As modeling and accurately predicting the AGN and XRB X-ray emission from the simulations is complex, we defer a thorough investigation of this discrepancy to future studies, where we will also replicate the whole analysis pipeline from mock simulation data, including the masking process.}
%
%
%
%
%
\section{Discussion}
\label{sec:discussion}

The combination of eROSITA's stable background and good sensitivity at moderate spatial resolution with the availability of a highly complete spectroscopic galaxy sample from GAMA allowed us to detect the faint X-ray emission around galaxies as a function of their measured stellar masses and sSFRs.

The work presented here shows a clear dichotomy in the average X-ray emission of star-forming and quiescent galaxies. While the former are only significantly detected on small scales, with a projected luminosity profile consistent with the eROSITA PSF and an intensity compatible with the faint end of the AGN population (with a possible contribution from XRBs), the latter show clearly extended projected emission, with increasing intensity for larger stellar masses (at least for $\log_{10} M^* > 11.2$). 

In this section, we first discuss possible systematic effects that could affect the interpretations of our results, ranging from the estimate of the contribution from undetected faint AGN or XRBs (Sect. \ref{subsubsec:agn_cont_discussion}) to the accuracy of the sample selection and the effect of the 2D projection of the large-scale emission surrounding the galaxies (Sect. \ref{subsec:disc:projection}). In Sect. \ref{subsec:disc:simu}, we move to a discussion of the comparison with the numerical simulation predictions.

\subsection{AGN and XRB contamination}
\label{subsubsec:agn_cont_discussion}

The procedure to mask the event files according to a given criterion has a significant impact on the results (see Sect. \ref{subsec:full:stack}). Using ALL, M1, or M3 masks leads to drastically different measurements with distinct physical meanings. 
In our case, when all resolved sources are identified and classified, the M1 mask is then optimal, as it corresponds to minimization of the contamination from known AGN to the inner projected luminosity profiles. However, the level of residual contamination from undetected faint AGN remains uncertain.

To further estimate the AGN contamination, we use the X-ray AGN model described in \citet{Comparat2019MNRAS.487.2005C} and validated against eFEDS observations by \citet{Liu_SIM_2021arXiv210614528L}. 
We use the X-ray group and cluster model from \citet{Comparat2020OJAp....3E..13C}, which was validated against eFEDs observations by \citet{LiuAng2021arXiv210614518L}, \citet{Bulbul2021arXiv211009544B}. 
In particular, the simulations accurately reproduce  the number density of sources as a function of their flux (logN-logS) for each class separately. 
We limit these light cones to the redshift range $0.05<z<0.3$ and flux in the band 0.5--2 keV to $F_X> 10^{-17}$ erg cm$^{-2}$ s$^{-1}$. 
Figure~\ref{fig:scaling:relation:EMP:model} shows a comparison of the observations with the empirical models. 
The observations are compatible with the two model lines: at low stellar masses ($\log_{10} M^* < 11.$) the total projected emission of both quiescent and star-forming galaxies must be contaminated by faint AGN\footnote{We note here that the AGN model prediction is in good agreement with the figures from Aird et al. (in prep) quoted in Sect. \ref{subsec:agn}.}. The corresponding contamination from unresolved emission due to XRBs based on the adopted model for the XRB population in galaxies \citep[see discussion in Sect. \ref{subsec:xrb};][]{Aird2017MNRAS.465.3390A} is reported in Table \ref{tab:LX:80kpc:M3} (M3 mask) and is shown as colored shaded areas in Fig.~\ref{fig:scaling:relation:EMP:model}. Clearly, the putative contribution from XRBs remains subdominant in our stacked subsamples.

While the comparison with the empirical models provides a reasonable explanation for the observed projected luminosity of both star-forming galaxies (in terms of faint AGN) and high-mass quiescent galaxies (in terms of virialized hot haloes), it does not satisfactorily explain the detection of a significant projected emission well beyond 80 kpc  around low-mass
quiescent galaxies, which is much more extended than the potential contamination from point sources alone (see Fig.~\ref{fig:psf:vs:redshift}). 
Additional sources of genuinely extended X-ray emission surrounding those lower mass quiescent galaxies are likely needed. These may be due to the CGM itself, to an incomplete removal of satellites of luminous clusters and groups, or to projection effects. We discuss these latter two possibilities below.

\subsection{Central versus satellites and projection effects}
\label{subsec:disc:projection}

A well-defined central galaxy sample is key to enabling the interpretation of our results. 
{Galaxies and their haloes are not generally isolated; they reside in clustered environments \citep{Mandelbaum2006MNRAS.368..715M, Gillis2013MNRAS.431.1439G}. For example, when measuring projected statistics, the signal coming from central galaxies with high clustering (living in a clustered, dense environment) will be boosted compared to that of isolated galaxies. This is because of neighboring central galaxies that are in projection within 300kpc but in three dimensions within a few Mpc.}
With its high completeness, namely of $\sim98\%$, the GAMA spectroscopic survey allows an accurate distinction to be made between central and satellite galaxies in different environments; 
the removal of satellite galaxies is then straightforward, and this procedure should not induce systematic effects provided the GAMA completeness is uniformly high irrespective of environment or galaxy properties. 
Conversely, we also carried out the same measurements in a sample where satellite galaxies are included and found that the SF profiles remained unchanged, while the QU profiles were systematically 25\% brighter at all scales. 
This is consistent with the fact that, on average, quiescent satellite galaxies live in hot and dense environments \citep[e.g.,][]{Velander2014MNRAS.437.2111V, Hudson2015MNRAS.447..298H}. 

\begin{figure}
    \centering
    \includegraphics[width=.95\columnwidth ]{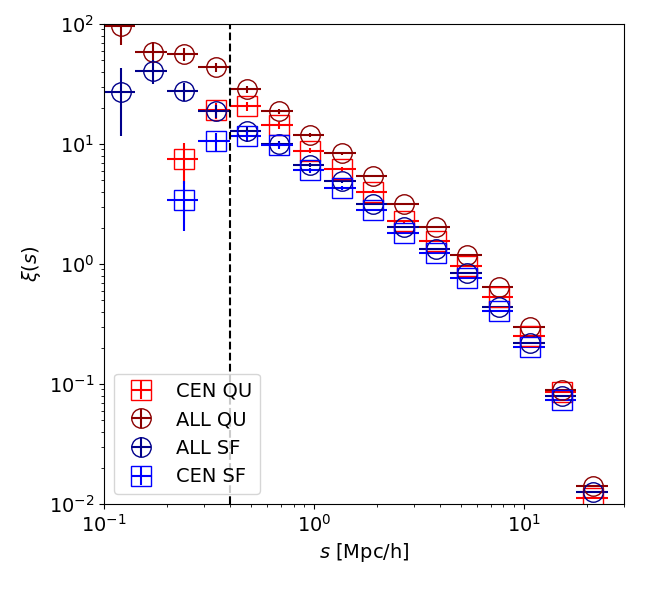}
    \includegraphics[width=.95\columnwidth ]{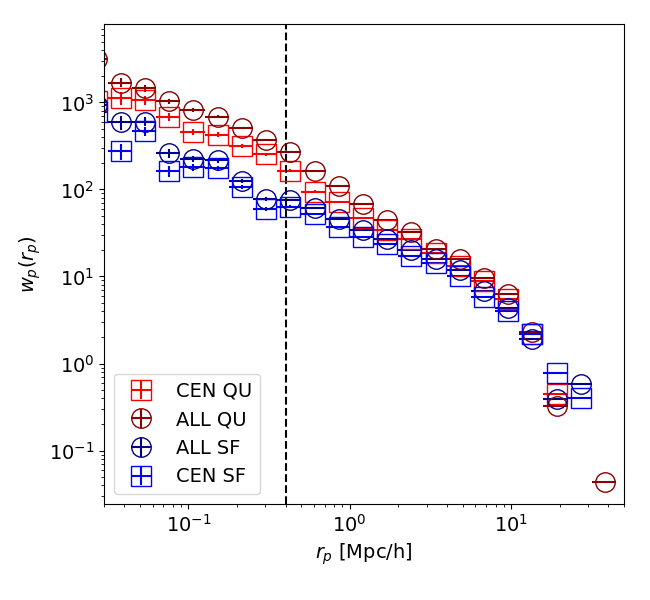}
    \caption{Clustering of the samples selected. In each panel we show the results for the quiescent central galaxies (CEN QU, red squares), the star-forming centrals (SF CEN, blue squares), all the quiescent galaxies (ALL QU, purple circles), and all the star-forming ones (ALL SF, dark blue circles). \textit{Top Panel:} Three-dimensional correlation function. The turn over  due to the central selection function is clear, as illustrated
by the vertical dashed line. \textit{Bottom Panel:} Projected correlation function: no turnover is visible.}
    \label{fig:clustering}
\end{figure}

{To illustrate the projection effects, we measured the clustering (two-point correlation function) of the galaxies considered here using two statistics: $w_p(r_p)$ and $\xi(s)$. $\xi(s)$ is the angle-averaged ("isotropic") 3D redshift space correlation function and $w_p(r_p)$ is the line-of-sight projected correlation function \citep[for detailed definitions see e.g.,][]{Davis1983ApJ...267..465D}.}
The luminosity profiles obtained are subject to projection, similarly to the $w_p(r_p)$ clustering statistics; the environment in which galaxies live has an effect on projected statistics. To investigate the impact of projection effects, we measured the clustering of the galaxy samples considered, before and after the central selection algorithm is applied; see Fig. \ref{fig:clustering}. 
In the 3D correlation function of the central galaxies sample ($\xi(s)$, top panel, red or blue squares), there is a clear cut-off scale ($\sim400$ kpc, identified by vertical dashes) below which the clustering signal diminishes, while it is still increasing in the complete sample (central plus satellite galaxies, red and blue circles, respectively). 
This means that the selection procedure for central galaxies works as expected.

When measuring the projected clustering $w_p(r_p)$, the cut-off scale at $\sim400$ kpc is no longer visible; see the bottom
panel of  Fig. \ref{fig:clustering}. The projected clustering of the full sample and the central sample both extend to very small scales. We clearly see a projection effect here, which is due to the fact that galaxies live in crowded, {clustered} environments. 
Therefore, when measuring the projected luminosity profiles, the emission around individual galaxies is indistinguishable from the emission from the environment of the galaxies. 
Both star-forming and quiescent galaxy samples are subject to a similar projection effect, as shown in Fig.~\ref{fig:clustering}. However,
from the difference seen between the star-forming and quiescent stacked profiles, we infer that the environment of $5\times10^{10}M_\odot$ (and lower) quiescent galaxies is dense and hot, while that of star-forming galaxies needs to be {either less dense, cooler, or both}. 
This is in agreement with the findings of \citet{Velander2014MNRAS.437.2111V}. 
Accurately quantifying  the strength of the projection effect and how much it biases the projected luminosity profile as a function of scale is key to determining how much of the measured emission comes from the CGM. 
Future simulations (with full fledged light cones) and further modeling in order to jointly interpret halo occupation distributions {(constrained with clustering and galaxy-galaxy lensing measurements)} and the luminosity profile will hopefully shed light on this uncertainty. 
Because {of this}, it is unclear whether or not the CGM around $5\times$10$^{10}$ M$_\odot$ galaxies is detected in our sample. 
The eROSITA eFEDS survey is not deep enough to provide a complete census of galaxy groups in the redshift range studied here, and so applying the ALL mask will remove only a large part of the environmental effects, but not all.

\subsection{Insights from the comparison between observations and simulation results}
\label{subsec:disc:simu}

The X-ray profiles and luminosities predicted at large galactocentric distances by TNG100 and EAGLE are in general agreement with observations, and so this allows us to use the simulations to also gather insights that are not obtainable otherwise.
It is important to point out that the simulated mean profiles are significantly brighter than the simulated median profiles: see Fig.~\ref{fig:full:stack:SF:QU:SIMS}, thick versus thin solid (TNG100) and dashed (EAGLE) curves. Conversely, this implies that the CGM signal of the median galaxy is not properly captured by the observed mean stacks, which are instead biased high. 
By analyzing the TNG100 results, we can quantify that the 10\%\ most luminous sources alone can bias high the X-ray signal at $\sim100$ kpc by almost 1 dex. 
This means that caution is necessary in the interpretation of the results. 
Chiefly, given the small volumes considered in this analysis, it is hard to tightly control the population of rare objects in the same manner in the observations and simulations. Samples collected over larger volumes will undoubtedly help in addressing this in the future.
In the meanwhile, we notice that in TNG100 and EAGLE, no matched galaxy that enters in this analysis resides in a halo more massive than $3\times10^{13}$--$10^{14}$ M$_\odot$, that is, throughout the whole sample and even though more massive haloes are present in the simulated volumes. On the other hand, there are at least 289 (X-ray-detected) galaxy clusters and groups in the 60 deg$^2$ of eFEDS and GAMA \citep{LiuAng2021arXiv210614518L}.

Also, when comparing results from the two simulations, TNG100 and EAGLE, we find overall very consistent mean profiles, whereas median profiles differ. 
The median discrepancy is related to the different underlying galaxy physics models of the simulations. 
This highlights the fact that the median profiles are sensitive to the galaxy evolution model, while the mean profiles are more sensitive to a sparse population of rare and luminous galaxies. 
For example, at the transitional mass regime of Fig.~\ref{fig:full:stack:SF:QU:SIMS}, it is evident that EAGLE atmospheres are brighter than TNG ones in X-rays, which is consistent with the notion that the SMBH feedback in TNG is more ejective than in EAGLE \citep{Davies2020,Truong2021b} and that EAGLE exhibits lower quiescent fractions than TNG at this mass scale \citep{Donnari2021}.

At face value, {in the 80-300kpc range}, the X-ray luminosity of the CGM around $10^{10.2-10.8}\,\MSUN$ star-forming galaxies is consistent with the upper limits inferred from eFEDS+GAMA data.
The simulation results suggest that a detection of the hot CGM around star-forming MW/M31-mass galaxies may soon be obtained (see Sect.~\ref{sec:summary}).

Importantly, it is also apparent from both Figs.~\ref{fig:full:stack:SF:QU:SIMS} and \ref{fig:scaling:relation:wSim} that TNG100 and EAGLE do not predict a significant difference in the CGM X-ray signals between quiescent and star-forming galaxies, as seen in the data. 
Instead, the mean profiles and the CGM luminosities on large scales from the simulations fall closer to the observed SF eROSITA results. 
This is likely due to the aforementioned projection effects. 

Also, it should be noted that these results are not necessarily in contradiction with results reported in Section 1 which suggest brighter X-ray atmospheres around simulated star-forming galaxies than around quiescent ones \citep{Truong20, Oppenheimer2020}. First, here we are comparing mean profiles and not median properties of galaxies; second, we focus on somewhat higher redshift, $z\sim 0.2$ versus $z\sim0.1$ {(1.1 Gyr difference)}; and finally, the stacked profiles of Figure~\ref{fig:full:stack:SF:QU:SIMS} are not from volume-limited samples and reflect the mean results of galaxies in rather wide mass and redshift bins and with different stellar mass and redshift distributions in the two star-forming and quiescent bins, with the former exhibiting a greater representation of lower mass and higher redshift galaxies (see Fig.~\ref{fig:subm1:107:comparison:hist:comparison}).

As pointed out above, for more accurate comparisons between observations and simulations, a more thorough forward model of the simulation data would be needed, for example using \texttt{sixte} \citep{Dauser2019AA630A66D}. 
However, focusing on the discrepancy at face value for the $5\times10^{10}$ M$_\odot$ quiescent galaxies (Fig. \ref{fig:full:stack:SF:QU:SIMS}), we can argue that under-luminous atmospheres in the simulations in comparison to data may indicate that (some of the) simulated gaseous haloes may (a) be under-dense, (b) be characterized by lower temperatures, and/or (c) be less enriched than (some of the) galaxies in the Universe, or more specifically in eFEDS. 
{This discrepancy} could be due to a SMBH feedback implementation that is more ejective than in reality, at least in some cases ---more ejective SMBH feedback would imply more substantial outflows and hence a more substantial clearing of the halo of both hot and metal-enriched gas--- or that is not effective enough at heating up the halo gas.

%
%
%
%
\section{Summary and outlook}
\label{sec:summary}
We quantified the X-ray emission around a large sample of quiescent (star-forming) galaxies at $0.05<z<0.3$ in the stellar mass range $2\times 10^{10}$ -- $10^{12}$ M$_\odot$ ($3\times 10^{9}$ -- $6\times10^{11}$ M$_\odot$) (Fig. \ref{fig:GAMA:data}). 
To do so, we stacked the eROSITA eFEDS events around central GAMA galaxies to obtain projected luminosity profiles out to hundreds of project kpc (Figs. \ref{fig:full:stack:measurement}, \ref{fig:full:stack:SF:QU}, and \ref{fig:subm1:trends}). 
As anticipated, the stacking method is successful in overcoming the flux limit in the X-ray observations
\footnote{For example, the QU\_M10.53 sample, with a S/N of about 5.5, has a luminosity in the inner 80 kpc of 3.2$\times 10^{40}$ erg s$^{-1}$ at redshift 0.19, corresponding to a flux of 3.1$\times 10^{-16}$ erg cm$^{-2}$ s$^{-1}$. 
This is about 20 times fainter than the flux limit of  quoted by \citet{Brunner2021arXiv210614517B}.}. 

For quiescent (passive) galaxies, the X-ray profiles are clearly extended throughout the available mass range (e.g., Fig. \ref{fig:full:stack:SF:QU}); however,
the measured profiles are likely biased high due to projection effects emanating from the fact that quiescent galaxies live in dense and hot environments (Fig. \ref{fig:clustering}).
Around star-forming galaxies with $< 10^{11}$ M$_\odot$, the X-ray-stacked profiles are compatible with unresolved sources and are consistent with the expected emission from faint AGN and XRBs (Fig. \ref{fig:full:stack:SF:QU}). Only for the most massive star forming samples ($\geq10^{11}$ M$_\odot$) is there a hint of detection of extended emission. 

We measure for the first time the average relation between mean projected X-ray luminosity within various apertures and stellar mass separately for quiescent and star-forming galaxies (Fig. \ref{fig:scaling:relation}). 
We find that the relation is different for the two galaxy populations: high-mass ($\geq 10^{11}$ M$_\odot$) star-forming or quiescent galaxies follow the expected scaling of virialized hot haloes ({see orange solid line in Fig. 7}). 
Lower mass star-forming galaxies show a less prominent luminosity that is also more weakly dependent on stellar mass, consistent with empirical models of the population of weak AGN.

In particular, when measuring the mean projected X-ray luminosity in a 300 kpc aperture while excluding X-ray-bright AGN detected as point sources (M1 mask, Fig. \ref{fig:scaling:relation} main panel and Table \ref{tab:LX:80kpc:M1}),  
for quiescent galaxies with a stellar mass larger than $2\times 10^{10}$ M$_\odot$, we detect (S/N larger than 3) a faint X-ray emission partly originating from hot gas.  
For star-forming galaxies with a stellar mass of larger than $6\times 10^{10}$ M$_\odot$, we report a hint of detection (S/N between 1 and 3). 
For star-forming galaxies with a stellar mass in the range $3\times 10^{9}$--$6\times 10^{10}$ M$_\odot$ we measure upper limits (S/N smaller than 1). We find similar results and detection levels when we measure the average projected X-ray luminosity in a 80--300 kpc aperture (M1 mask, Fig. \ref{fig:scaling:relation} bottom right panel and Table \ref{tab:LX:80kpc:M1}), which hence characterize the extended emission beyond the galaxy itself: we detect X-ray extended emission around quiescent galaxies at all probed masses, while for star-forming galaxies we find upper limits in the $3\times 10^{9}$--$10^{11}$ M$_\odot$ range and a hint of detection at higher masses.

We additionally measure the average projected X-ray luminosity in a 300 kpc aperture while keeping all detected point sources (M3 mask, Fig. \ref{fig:scaling:relation:EMP:model} and Table \ref{tab:LX:80kpc:M3}). 
We find good agreement with empirical models of the X-ray cosmic web of AGN and clusters from \citet{Comparat2019MNRAS.487.2005C, Comparat2020OJAp....3E..13C}. This reinforces the notion that a robust assessment of the properties of hot haloes in Milky-Way-like galaxies (and smaller) requires accurate removal of weak AGN contaminants.

When comparing our results with state-of-the-art numerical simulations (IllustrisTNG and EAGLE), we find overall consistency in the average emission at large ($>80$ kpc) scales and at masses $\geq 10^{11}$ M$_\odot$, but disagreement at smaller scales, where brighter-than-observed compact cores are predicted (Fig. \ref{fig:full:stack:SF:QU:SIMS}). 
The simulations also do not predict the clear differentiation that we observe between quiescent and star-forming galaxies in our samples (Fig. \ref{fig:scaling:relation:wSim}).

This work is a stepping stone towards a better understanding of the hot phase in the CGM, which holds a key role in the regulation of star formation.
In the next decade, by combining eROSITA with upcoming spectroscopic galaxy surveys \citep[e.g., DESI, SDSS-V, 4MOST:][]{DESI2016, Kollmeier2017arXiv171103234K, Jong2019Msngr.175....3D, Merloni2019Msngr.175...42M, Finoguenov2019Msngr.175...39F}, properties of the CGM and their relation to AGN should be unraveled. 
The eROSITA all-sky survey (eRASS) data cover more than 100 times more extragalactic sky than eFEDS and therefore offer the opportunity to improve on the analysis presented here. 

Table~\ref{tab:erass}  provides a rough estimate of the average S/N improvement ratio ($\xi_{\overline{\rm S/N}}$) based on a full-sample for a few selected combinations of different eRASS depths (from the single-pass eRASS1 to the full eight-pass eRASS:8) and extragalactic spectroscopic surveys. 
We assume, for simplicity, that $\overline{\rm S/N}\propto \sqrt{N_g t_{\rm exp}}$, where $N_g$ is the number of galaxies in the range $z\lesssim 0.3$ and $t_{\rm exp}$ is the average eROSITA exposure. 
We compute the ratio of the S/N, $\xi_{\overline{\rm S/N}}$, obtainable with these experiments to the eFEDS+GAMA09 baseline one, for which we take $N_{g,eFEDS}=$35,521.
We provide estimates for the following combinations:
\begin{itemize}
    \item eRASS1 with a compilation of existing spectroscopic catalogs (`eRASS1 + public 2021');
    \item An intermediate stage that combines a three-pass X-ray survey (eRASS:3) with the DESI `Bright Galaxy Survey' \citep[BGS,][]{Ruiz2020RNAAS...4..187R}; 
    \item The full-depth eRASS:8 combined with the 4MOST WAVES wide survey \citep{Driver2016ASSP...42..205D};
    \item A putative `Legacy' sample, combining eRASS:8 with all spectroscopic samples available in about a decade (early 2030s).
\end{itemize}

\begin{table}
    \centering
    \caption{\label{tab:erass} Forecast  S/N for future eROSITA-based experiments.} 
    \begin{tabular}{c | c c c c }
\hline \hline
& Area  & $N_g$  & & \\
& [deg$^2$] & [10$^6$] & $\langle \frac{t_{\rm exp}} {t_{\rm exp, eFEDS}} \rangle$  & $\xi_{\overline{\rm S/N}}$ \\
\hline
eRASS1 + public 2021 & 21k & 1.2 & 0.1 & 1.8 \\
eRASS:8 + WAVES & 1k & 0.7 & 0.8 & 4.0 \\
eRASS:3 + DESI BGS & 5k & 7 & 0.3 & 7.7 \\
eRASS:8 `Legacy' & 21k & 20 & 0.8 & 21.2 \\

\hline
    \end{tabular}
\tablefoot{Average S/N improvement ratio with respect to the eFEDS+GAMA09 sample ($\xi_{\overline{\rm SNR}}$) for four possible combinations of eROSITA all-sky survey depths and low-redshift galaxy spectroscopic samples. The Area (in square degrees) reported is the approximate overlap between the German eROSITA all-sky survey \citep{Sunyaev2021arXiv210413267S} and the galaxy samples; $N_g$ is the number of galaxies with $z\lesssim 0.3$ in this area (in millions), while $\langle \frac{t_{\rm exp}} {t_{\rm exp, eFEDS}} \rangle$ is the average ratio between the corresponding eRASS exposure and the eFEDS one.}    
\end{table}

The average S/N improvement with respect to the sample analyzed here ranges from about a factor of 1.8 to more than a factor of 20.
Further improvements in the eROSITA data processing, energy, and PSF calibration will likely also increase the significance of the detection, and enable a combination of spatial and spectral analysis. Thanks to those improvements, we should be able to measure the temperature and metallicity of the hot CGM, as well as its density and pressure profile. 
Hopefully, this will shed light on how hot haloes are created and energized, and their interplay with the star formation and virialization processes, as well as feedback processes from AGN. 


\begin{acknowledgements}
This work is based on data from eROSITA, the soft X-ray instrument aboard \textit{SRG}, a joint Russian-German science mission supported by the Russian Space Agency (Roskosmos), in the interests of the Russian Academy of Sciences represented by its Space Research Institute (IKI), and the Deutsches Zentrum f\"ur Luft- und Raumfahrt (DLR). The \textit{SRG} spacecraft was built by Lavochkin Association (NPOL) and its subcontractors, and is operated by NPOL with support from the Max Planck Institute for Extraterrestrial Physics (MPE).

The development and construction of the eROSITA X-ray instrument was led by MPE, with contributions from the Dr. Karl Remeis Observatory Bamberg \& ECAP (FAU Erlangen-N\"urnberg), the University of Hamburg Observatory, the Leibniz Institute for Astrophysics Potsdam (AIP), and the Institute for Astronomy and Astrophysics of the University of T\"ubingen, with the support of DLR and the Max Planck Society. The Argelander Institute for Astronomy of the University of Bonn and the Ludwig Maximilians Universit\"at Munich also participated in the science preparation for eROSITA. The eROSITA data shown here were processed using the eSASS software system developed by the German eROSITA consortium.

GAMA is a joint European-Australasian project based around a spectroscopic campaign using the Anglo-Australian Telescope. The GAMA input catalog is based on data taken from the Sloan Digital Sky Survey and the UKIRT Infrared Deep Sky Survey. Complementary imaging of the GAMA regions is being obtained by a number of independent survey programmes including GALEX MIS, VST KiDS, VISTA VIKING, WISE, Herschel-ATLAS, GMRT and ASKAP providing UV to radio coverage. GAMA is funded by the STFC (UK), the ARC (Australia), the AAO, and the participating institutions. The GAMA website is http://www.gama-survey.org/ .

This project acknowledges funding from the European
Research Council (ERC) under the European Union’s Horizon
2020 research and innovation programme (grant agreement No
865637).

\end{acknowledgements}

\bibliographystyle{aa}
\bibliography{references}

\clearpage
\begin{appendix}
\section{Matching procedure between simulated and observed galaxy samples}
\label{app:matching}

To directly compare the results of the simulations for example from \citealt{Truong20} and \citealt{Oppenheimer2020} to those from our observations is a complex undertaking, first and foremost because they are given at different redshifts ($z\sim0$ vs. median $z\sim0.2$), which correspond to a 2.5 Gyr difference in the age of the Universe. 
Second, the galaxy samples from the simulations at any given epoch are volume limited whereas those in GAMA are magnitude limited, with progressively larger fractions of more luminous and massive galaxies at higher redshifts (see Figure\ref{fig:GAMA:data}). 

We therefore constructed simulated galaxy samples that are matched to the observed ones by finding, for each galaxy in the GAMA set, its simulated equivalent in TNG and EAGLE. 
We used simulated data from the TNG100 and Ref-L0100N1504 runs, which both encompass approximately 100 comoving Mpc a side. 
We consider only central galaxies, in line with the choices of Sect. \ref{subsec:centrals}. However, we do not replicate the methods adopted for the observed data, but simply identify as centrals those galaxies that occupy the lowest level of the gravitational potential of the simulated haloes identified with the FoF algorithm. In TNG, we exclude objects with SubhaloFlag $\equiv0$ \citep[see][for more details]{nelson2019a}.
The analogous objects between the simulation outputs and the GAMA sample are found in the redshift, galaxy stellar mass, and sSFR parameter space. 

The match in redshift is done to the closest available simulation snapshot. In the $0.05<z<0.3$ range considered here, there are 18 data snapshots in TNG and 3 in EAGLE. As a fiducial choice, we assume that the stellar mass of GAMA galaxies inferred by \citet{Bellstedt2020MNRAS.498.5581B, Bellstedt2021MNRAS.503.3309B} is close to the total mass of a galaxy, i.e., is compatible with the sum of all the stellar particles that are gravitationally bound to a simulated galaxy. However, this is probably not the case for centrals in massive haloes and so we bracket this uncertainty by also matching the samples assuming that the GAMA stellar masses correspond to the sum of the stellar particles that are gravitationally bound to a simulated galaxy and within twice its stellar half-mass radius.

We also assume that the inferred SFR of the observed galaxies is equivalent to the instantaneous SFR of the gas cells of a simulated galaxy, again within twice its stellar half-mass radius. 
In fact, GAMA SFRs are averaged over 100 Myr: as shown by \citealt{Donnari2021}, at least for TNG, whether SFRs are de facto instantaneous or averaged across the past 1 Gyr and whether they are measured across varying galaxy apertures should not be important for the main purposes of this paper, that is, for splitting galaxies into quiescent and star-forming ones at low $z$ and as long as the rule for the grouping is the same; see below. 
However, even within the quiescent and star-forming samples, the X-ray CGM signal may show trends with the actual level of SFR, and hence we also show results adopting 100 Myr-averaged values measured throughout the galaxy body. We manually re-label sSFR values below $10^{-15}$ yr$^{-1}$ to exactly $10^{-15}$ yr$^{-1}$ in both observation and simulations data in order to avoid being affected by highly uncertain or numerical resolution-dependent SFR estimations.

For every GAMA galaxy at a given redshift (see Table~\ref{tab:gama:data}), a simulated analog is searched for in the stellar mass--sSFR plane at the closest snapshot by randomly selecting a galaxy within a rectangle whose widths are equivalent to the uncertainties associated to the measured values of stellar mass and SFR on a galaxy-by-galaxy basis. If no simulated galaxy is found within such limits, we simply take the closest simulated galaxy to the observed one; this occurs only for $\sim400$ ($\sim800$) GAMA galaxies when matching to TNG (EAGLE). 
We repeat this procedure for the overall sample 20 times for both TNG and EAGLE so as to obtain 20 different Monte Carlo simulation samples matched to the observational ones. 

Quiescent and star-forming matched simulated galaxies are then divided with the fixed boundary cut at $\log_{10}(\rm{sSFR})=-11$, as in Sect. \ref{sec:GAMA}. Bins in galaxy stellar mass as described in Table~\ref{tab:gama:data} are extracted from the overall matched samples.

As the TNG and EAGLE volumes ($\sim10^6$ Mpc$^3$) are smaller than that of the galaxy sample considered here ( $\sim10^7$ Mpc$^3$), 
the same simulated galaxies may be matched to multiple observed galaxies. 
In these cases, the direction from which the galaxy is seen (when projected on sky) is changed to avoid repetitions of the exact same X-ray profiles when simulating the stacking procedure. 
In the matched TNG100 samples, about $30\%$ of the objects come from simulated galaxies that are not unique. This fraction increases to $43\%$ for galaxies M$^*>10^{11}$ M$_\odot$. {The nonunique fraction is significantly higher for the EAGLE simulation ($\sim75\%$) because it has fewer snapshots (only 3) in the observed $0.05<z<0.3$ range. Therefore, on average, an EAGLE galaxy could be matched to four or five observed galaxies. This could be an issue in the case of outliers, that is, particularly bright systems, such as merging galaxies, because in such cases their contribution to the predicted mean average X-ray emission is artificially overestimated by many factors because of the duplication in the matching. In this regard, we noticed a merging galaxy in EAGLE in the mass bin $M_*\sim10^{10.7}M_\odot$ that would have been matched in different projections to five different observed galaxies, causing the EAGLE mean surface brightness to be about one order of magnitude higher in the central regions in comparison to the case when the system is omitted. Hence, to minimize the effect of the matching duplication in the fiducial EAGLE results presented throughout, we opt to count the contribution of that merging galaxy only once.}
\end{appendix}

\end{document}